\documentclass[twocolumn]{aastex631}

\usepackage{mhchem}
\usepackage{subfigure}

\newcommand{\logX}[1]{\ensuremath{\log(\mathrm{X_{\ce{#1}}})}}
\newcommand{\logXratio}[2]{\ensuremath{\log(\mathrm{X_{\ce{#1}} / X_{\ce{#2}} })}}

\begin{document}

\title{JWST/NIRISS reveals the water-rich ``steam world'' atmosphere of GJ 9827 d}

\newcommand{\umontreal}{Department of Physics and Trottier Institute for Research on Exoplanets, Universit\'{e} de Montr\'{e}al, Montreal, QC, Canada \href{mailto:caroline.piaulet@umontreal.ca}{caroline.piaulet@umontreal.ca}}

\author[0000-0002-2875-917X]{Caroline Piaulet-Ghorayeb}
\affil{\umontreal}

\author[0000-0001-5578-1498]{Bj\"{o}rn Benneke} 
\affil{\umontreal}

\author[0000-0002-3328-1203]{Michael Radica} 
\affil{\umontreal}

\author[0009-0002-2380-6683]{Eshan Raul}
\affil{Department of Astronomy, University of Michigan, Ann Arbor, MI, USA}

\author[0000-0002-2195-735X]{Louis-Philippe Coulombe} 
\affil{\umontreal}

\author[0000-0003-0973-8426]{Eva-Maria Ahrer}
\affil{Max Planck Institute for Astronomy, K\"{o}nigstuhl 17, D-69117 Heidelberg, Germany}

\author[0000-0001-9137-9818]{Daria Kubyshkina}
\affil{Space Research Institute, Austrian Academy of Sciences, Schmiedlstrasse 6, 8042 Graz, Austria}

\author[0000-0002-0583-0949]{Ward S. Howard}
\affil{Department of Astrophysical and Planetary Sciences, University of Colorado, 2000 Colorado Avenue, Boulder, CO 80309, USA}

\author[0000-0001-6878-4866]{Joshua Krissansen-Totton}
\affil{Department of Earth and Space Sciences/Astrobiology Program, University of Washington, Seattle, WA, USA}

\author[0000-0003-4816-3469]{Ryan MacDonald}
\affil{Department of Astronomy, University of Michigan, Ann Arbor, MI, USA}

\author[0000-0001-6809-3520]{Pierre-Alexis Roy} 
\affil{\umontreal}

\author[0000-0002-3191-2200]{Amy Louca}
\affil{Leiden Observatory, Leiden University, P.O. Box 9513, 2300 RA Leiden, The Netherlands}
\affil{SRON Netherlands Institute for Space Research, Niels Bohrweg 4, 2333 CA Leiden, The Netherlands}

 \author[0000-0002-4997-0847]{Duncan Christie}
\affil{Max Planck Institute for Astronomy, K\"{o}nigstuhl 17, D-69117 Heidelberg, Germany}

\author[0000-0002-5428-0453]{Marylou Fournier-Tondreau}
\affil{University of Oxford, Department of Physics Oxford, OX1 3PW, UK}

\author[0000-0002-1199-9759]{Romain Allart}
\affil{\umontreal}

\author[0000-0002-0747-8862]{Yamila Miguel}
\affil{Leiden Observatory, Leiden University, P.O. Box 9513, 2300 RA Leiden, The Netherlands}
\affil{SRON Netherlands Institute for Space Research, Niels Bohrweg 4, 2333 CA Leiden, The Netherlands}

\author[0000-0002-0298-8089]{Hilke E. Schlichting}
\affil{Department of Earth, Planetary, and Space Sciences, University of California, Los Angeles, Los Angeles, CA 90095, USA}

\author[0000-0003-0156-4564]{Luis Welbanks}
\affil{School of Earth \& Space Exploration, Arizona State University, Tempe, AZ 85257, USA}

\author[0000-0001-9291-5555]{Charles Cadieux}
\affil{\umontreal}

\author[0000-0001-6110-4610]{Caroline Dorn}
\affil{Department of Physics and Institute for Particle Physics and Astrophysics, ETH Zurich, CH-8093 Zurich, Switzerland}

\author[0000-0001-5442-1300]{Thomas M. Evans-Soma}
\affiliation{School of Information and Physical Sciences, University of Newcastle, Callaghan, NSW, Australia}
\affiliation{Max Planck Institute for Astronomy, K\"{o}nigstuhl 17, D-69117 Heidelberg, Germany}

\author[0000-0002-9843-4354]{Jonathan J. Fortney}
\affil{Department of Astronomy \& Astrophysics, University of California, Santa Cruz, CA 95064, USA}

\author[0000-0002-5887-1197]{Raymond Pierrehumbert}
\affil{University of Oxford, Department of Physics Oxford, OX1 3PW, UK}

\author[0000-0002-6780-4252]{David Lafrenière}
\affil{\umontreal}

\author[0000-0002-9147-7925]{Lorena Acu\~{n}a}
\affil{Max Planck Institute for Astronomy, K\"{o}nigstuhl 17, D-69117 Heidelberg, Germany}

\author[0000-0002-9258-5311]{Thaddeus Komacek}
\affil{Department of Astronomy, University of Maryland, College Park, MD 20742, USA}

\author[0000-0001-5271-0635]{Hamish Innes}
\affil{Department of Earth Sciences, Freie Universit\"{a}t Berlin, Malteserstr. 74-100, 12249 Berlin, Germany}
\affil{Institute of Planetary Research, German Aerospace Center (DLR), Rutherfordstraße 2, 12489 Berlin, Germany}

\author[0000-0002-9539-4203]{Thomas G. Beatty}
\affiliation{Department of Astronomy, University of Wisconsin--Madison, Madison, WI 53706, USA}

\author[0000-0001-5383-9393]{Ryan Cloutier}
\affil{Department of Physics \& Astronomy, McMaster University, 1280 Main Street W, Hamilton, ON L8S 4L8, Canada}

\author[0000-0001-5485-4675]{René Doyon}
\affil{\umontreal}

\author[0009-0003-2576-9422]{Anna Gagnebin}
\affil{Department of Astronomy \& Astrophysics, University of California, Santa Cruz, CA 95064, USA}

\author[0009-0007-9356-8576]{Cyril Gapp}
\affil{Max Planck Institute for Astronomy, K\"{o}nigstuhl 17, D-69117 Heidelberg, Germany}

\author[0000-0002-5375-4725]{Heather A. Knutson}
\affil{Division of Geological and Planetary Sciences, California Institute of Technology, Pasadena, CA 91125, USA}





\begin{abstract}

With sizable volatile envelopes but smaller radii than the solar system ice giants, sub-Neptunes have been revealed as one of the most common types of planet in the galaxy. While the spectroscopic characterization of larger sub-Neptunes (2.5--4R$_\oplus$) has revealed hydrogen-dominated atmospheres, smaller sub-Neptunes (1.6--2.5R$_\oplus$) could either host thin, rapidly evaporating hydrogen-rich atmospheres or be stable metal-rich “water worlds” with high mean molecular weight atmospheres and a fundamentally different formation and evolutionary history. Here, we present the 0.6--2.8$\mu$m JWST NIRISS/SOSS transmission spectrum of GJ 9827 d, the smallest (1.98 R$_\oplus$) warm (T$_\mathrm{eq, A_B=0.3} \sim 620$\,K) sub-Neptune where atmospheric absorbers have been detected to date. Our two transit observations with NIRISS/SOSS, combined with the existing \textit{HST}/WFC3 spectrum, enable us to break the clouds-metallicity degeneracy. We detect water in a highly metal-enriched ``steam world'' atmosphere (O/H of $\sim 4$ by mass and H$_2$O found to be the background gas with a volume mixing ratio (VMR) of $>31$\%). We further show that these results are robust to stellar contamination through the transit light source effect. We do not detect escaping metastable He, which, combined with previous nondetections of escaping He and H, supports the steam atmosphere scenario. In water-rich atmospheres, hydrogen loss driven by water photolysis happens predominantly in the ionized form which eludes observational constraints. We also detect several flares in the NIRISS/SOSS light-curves with far-UV energies of the order of 10$^{30}$ erg, highlighting the active nature of the star. Further atmospheric characterization of GJ 9827 d probing carbon or sulfur species could reveal the origin of its high metal enrichment.

\end{abstract}




\section{Introduction} \label{sec:intro}


\subsection{The radius valley}
One of the most fundamental observational results from the study of exoplanets has been the discovery that small, close-in planets are bifurcated into two seemingly separate populations \citep{fulton_california-kepler_2017}. The small-planet size distribution displays a feature known as the ``radius valley'': an observed dearth of close-in small planets with sizes of 1.8 to 2.0 Earth radii around FGK stars. This statistical feature has been proposed to be a result of photoevaporation \citep{owen_evaporation_2017} where the mass loss is driven by the high-energy irradiation from the host star, core-powered mass loss \citep{ginzburg_core-powered_2018,gupta_signatures_2020} whereby hydrogen envelopes are lost as a result of the bolometric luminosity from the host star and the residual internal heat which is released from the planetary core as it cools down over time; or even the superposition of gas-rich and gas-poor planets that form before and after disk dissipation \citep{lee_primordial_2021,lopez_how_2018,zeng_growth_2019}. Recent studies suggest that for lower-mass stars, the gap in the radius distribution becomes more and more populated, which suggests enhanced diversity in the bulk volatile content of small planets orbiting low-mass stars \citep{luque_density_2022,ho_shallower_2024,venturini_fading_2024}.

\subsection{What masses tell us about compositions}
In each of these theoretical frameworks, the abundant ``sub-Neptunes'' comprising the $\gtrsim 2 R_\oplus$ population are low-density planets with cores made of a mix of ices and rocky material underlying a hydrogen-dominated envelope. Meanwhile, the smaller ``super-Earths'' are understood as rocky in composition, either due to their formation or as a result of atmosphere stripping. The precisely measured bulk densities of the smallest known ($\lesssim1.6 R_\oplus$) super-Earths agree well with rock/iron core compositions \citep{parc_super-earths_2024}, in line with theoretical expectations. The low densities of sub-Neptunes, on the other hand, can be equally well explained by a wider range of compositions with different ratios of H$_2$ compared to high mean molecular weight (HMMW) volatiles. Plausible compositions compatible with sub-Neptune densities span the continuum from the canonical view of percent-by-mass H/He envelopes of ``gas dwarfs'' \citep{fortney_framework_2013, benneke_jwst_2024}, to ``metal-rich'' or ``steam world'' atmospheres rich in HMMW volatiles amounting to tens of percent of the planet mass (Figure \ref{fig:mass_radius}, see also \citealt{aguichine_mass-radius_2021}, \citealt{piaulet_evidence_2023}).

Furthermore, recent theoretical \citep{madhusudhan_habitability_2021, innes_runaway_2023,dorn_hidden_2021} and observational findings \citep{madhusudhan_carbon-bearing_2023, holmberg_possible_2024,benneke_jwst_2024,luque_density_2022} support greater compositional diversity within the sub-Neptune population. While the coolest planets may be ``Hycean worlds'' with water in condensed liquid/ice phases under a H/He atmosphere, the warmer temperature of the vast majority of planets would instead imply a supercritical state for the water/volatile layer. Contrary to liquid water, supercritical water is highly miscible with hydrogen. Such warm, volatile-rich planets may therefore host exposed ``mixed'' envelopes where both H$_2$ and potentially large amounts of HMMW volatiles coexist \citep{wogan_jwst_2024, burn_radius_2024,benneke_jwst_2024}. In addition, water is expected to partition between the planetary envelope, the molten or solid mantle, and the metal core, leading to more uncertainties in planetary structure models \citep{schlichting_chemical_2021,luo_majority_2024,dorn_hidden_2021}.

Mass and radius measurements for individual planets are insufficient to resolve this compositional ambiguity due to an inherent degeneracy in the interior models \citep{rogers_framework_2010} which prevents the disentangling of thin unstable hydrogen layers from deep layers of HMMW volatiles. Moreover, this question cannot yet be resolved at the population level \citep{rogers_conclusive_2023} when accounting for the effects of both thermal evolution and photoevaporation.

\subsection{Exploring the diverse population of small sub-Neptunes}
\begin{figure}
    \centering
    \includegraphics[width=0.48\textwidth]{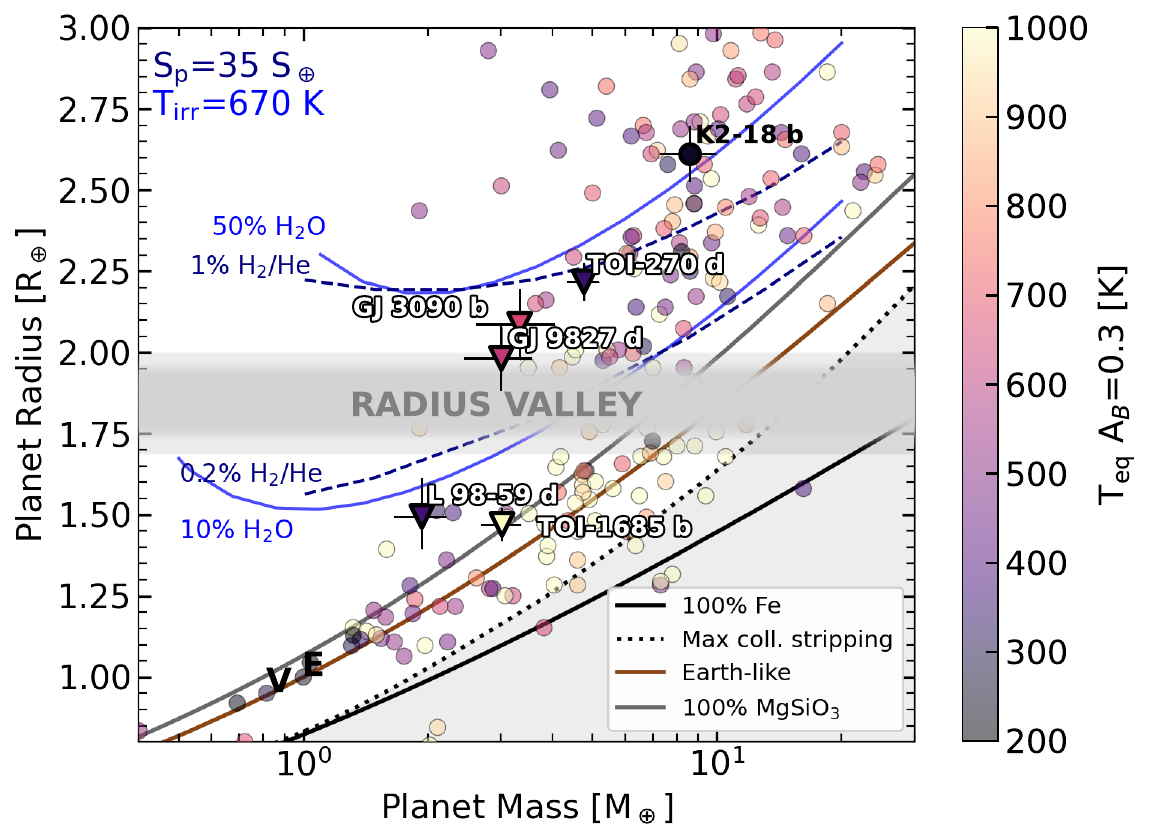}
    \vspace{5mm}
    \vspace{-5mm}\caption{Mass-radius diagram showing the small planets in the sample of JWST Cycle 2 Program GO 4098 (triangle markers) and how they compare to the larger K2-18 b. We also display planets from the NASA Exoplanet Archive with 3$\sigma$ or better mass measurements, colored by their equilibrium temperatures for an Earth-like albedo of 0.3. The planets span both sides of the radius valley (gray). We highlight the compositional degeneracy in mass-radius space with constant composition lines calculated using \texttt{smint} for planets with H$_2$/He envelopes or H$_2$O in the supercritical+vapor state (mass-radius relationships from \citet{lopez_understanding_2014,aguichine_mass-radius_2021}). We display mass-radius curves for water-rich compositions of 10\% and 50\% water in supercritical+vapor state assuming an irradiation temperature of 670 K (representative of GJ 9827 d; blue solid lines), and isocomposition curves for 0.2\% and 1\%-by-mass solar-composition H$_2$-dominated gas for a 1 Gyr planet receiving like GJ 9827 d approximately 35 times the Earth's insolation (navy dashed lines), both atop Earth-like cores. The positions of Venus and the Earth are indicated with black letters, and the light gray shaded area in the bottom right corner highlights planets smaller than expected from collisional stripping of planetary mantles \citep{marcus_minimum_2010}. With a mass of 3.02 $M_\oplus$ and a radius of 1.98 $R_\oplus$, GJ 9827 d hugs the upper size end of the radius valley, in line with its high predicted hydrogen escape rates.}
    \label{fig:mass_radius}
\end{figure}

Spectroscopic observations of planetary atmospheres can resolve these mass-radius degeneracies by providing precise measurements of the mean molecular weight in the upper atmosphere, modulo our understanding of how the amount of volatiles in the atmosphere translates to a bulk metal content \citep{thorngren_mass-metallicity_2016}. While atmospheric probes of large ($\gtrsim 2.4 R_\oplus$) Neptunes and sub-Neptunes revealed low atmospheric metallicities \citep{benneke_sub-neptune_2019,benneke_water_2019,madhusudhan_carbon-bearing_2023} in line with the expectations for a H$_2$/He-dominated atmosphere, more diversity is expected in the direct vicinity of the radius valley, as H$_2$/He atmospheres on smaller sub-Neptunes have greater susceptibility to atmospheric mass loss. We sought to explore this diversity of small sub-Neptune atmospheres through a large JWST Cycle 2 program (GO 4098), by obtaining high-signal-to-noise ratio (S/N) transmission spectra of small, low-density planets with well-characterized masses and radii (Figure \ref{fig:mass_radius}) across the 0.6$-$5.5$\mu$m range with JWST's Single Object Slitless Spectroscopy (SOSS) mode of the Near Infrared Imager and Slitless Spectrograph (NIRISS) instrument (NIRISS/SOSS) and with the G395H grism of the Near Infrared Spectrograph (NIRSpec/G395H). Because of the small sizes and low equilibrium temperatures of the most promising candidates, transmission spectroscopy is the method of choice to characterize their atmospheric compositions.

\subsection{Introduction to the GJ 9827 system}


GJ 9827 d is a 1.98 R$_\oplus$ planet transiting a moderately active nearby (30 pc), bright (J magnitude of $7.984 \pm 0.020$) K7V dwarf \citep{prieto-arranz_mass_2018}. It is one of three planets in a 1:3:5 near-resonant system including two other closer-in super-Earths, GJ 9827 b (1.57 R$_\oplus$) and GJ 9827 c (1.24 R$_\oplus$), detected by the \textit{K2} survey \citep{rodriguez_system_2018,niraula_three_2017}. The coevolution of these three planets spanning the radius valley under the irradiation of the same host star makes GJ 9827 an ideal system to test our understanding of how atmosphere escape shapes planetary atmospheres and compositions.

Radial velocity observations with the Magellan II Planet Finder Spectrograph \citep{teske_magellanpfs_2018}, the FIES, HARPS, and HARPS-N spectrographs \citep{prieto-arranz_mass_2018,rice_masses_2019}, supported by later ESPRESSO observations \citep{passegger_compact_2024} revealed that while GJ 9827 b and c have rocky densities, GJ 9827 d has a lower density consistent with a volatile envelope. Indeed, even when the joint evolution of all three planets is taken into account, the absence of an atmosphere on GJ 9827 b and c does not preclude the presence of an atmosphere on GJ 9827 d \citep{owen_testing_2020,kosiarek_physical_2021,bonomo_cold_2023}. 

The mass and radius of GJ 9827d are consistent with compositions ranging from a fraction of a percent H$_2$/He \citep{kosiarek_physical_2021,passegger_compact_2024}, or a mass fraction of HMMW volatiles of 5 to 30\% with a mixed-supercritical envelope state for this $T_\mathrm{eq, A_B=0.3}\simeq 620\,K$ planet \citep{aguichine_mass-radius_2021, innes_runaway_2023, benneke_jwst_2024}. Atmospheric escape considerations, however, disfavor a hydrogen-dominated atmosphere, as the intense mass loss predicted in this case from both photoevaporation \citep{carleo_multiwavelength_2021}, and core-powered mass loss \citep{gupta_caught_2021}, was faced with nondetections of Ly-$\alpha$, H$\alpha$ and He I using high-resolution space-based and ground-based observations \citep{kasper_non-detection_2020,carleo_multiwavelength_2021,krishnamurthy_absence_2023}.



While clouds often plague transmission spectra of sub-Neptune-size planets \citep{kreidberg_clouds_2014,wallack_jwst_2024}, the recently published \textit{HST}/WFC3 transmission spectrum of GJ 9827 d exhibits an absorption feature at 1.4 $\mu$m \citep{roy_water_2023} which suggests that even if clouds partially mute the transmission spectrum, molecular features can still be detected. The analysis of the HST spectrum favored water over methane (which has a coinciding absorption band at 1.4 $\mu$m, see \citealt{benneke_water_2019,madhusudhan_carbon-bearing_2023}) as the source of the absorption feature, mainly from the shape of the wings of the molecular band. However, a broader transmission spectrum that would probe multiple rotational/vibrational bands of both molecules could disentangle their relative contributions to the observed absorption features in transmission \citep{benneke_jwst_2024}. Photospheric heterogeneities in the K7-type host GJ 9827 are not expected to create significant contamination via the transit light source effect \citep{rackham_transit_2018}, lifting the ambiguity of the atmospheric or stellar origin of potential water absorption features \citep{roy_water_2023}.

Here we present the first JWST look at the atmosphere of GJ 9827 d using two NIRISS/SOSS transits. These observations are part of a large program that will additionally provide two transit observations with NIRSpec/G395H over the coming year. NIRISS/SOSS is sensitive to H$_2$O and CH$_4$ absorption, with multiple absorption bands of each molecule across the 0.6$-$2.8$\mu$m range. Furthermore, the resolution of NIRISS/SOSS opens up the possibility of constraining potential absorption from metastable He at 1083 nm \citep{fu_water_2022}. 

Our study is outlined as follows. We present the NIRISS/SOSS observations and data reduction in Section \ref{sec:obs}, followed by the white- and spectroscopic light-curve fitting procedure in Section \ref{sec:lc_analysis}. Our modeling of the atmospheric and stellar contributions to the spectrum is detailed in Section \ref{sec:atm_modeling}. We describe our atmospheric escape modeling in Section \ref{sec:escape_modeling}, and our planet structure modeling in Section \ref{sec:structure_modeling}. We present our results in Section \ref{sec:atm_results}, discuss them in Section \ref{sec:discussion}, and summarize our conclusions in Section \ref{sec:conclusion}.




\section{Observations and Data Reduction}\label{sec:obs}

In this study, we analyze two NIRISS/SOSS transit observations of GJ 9827 d using two independent data reduction pipelines. Throughout this study, we treat the \texttt{supreme-SPOON} reduction as our baseline and check that our results are not affected by the choice of reduction pipeline by comparing our transmission spectrum with those obtained from an independent \texttt{NAMELESS} reduction.

\subsection{Summary of the observations}\label{sec:niriss_obs}

Two transits of GJ 9827d were observed with JWST's SOSS mode \citep{albert_near_2023} of the NIRISS instrument \citep{doyon_near_2023} as part of the Cycle 2 General Observer program GO 4098 (PI: Benneke; Co-PI: Evans-Soma). Each NIRISS/SOSS time series observation (TSO) lasted 3.4 hours. We chose the SUBSTRIP256 subarray which enables us to extract the stellar spectrum from both order 1 (0.8 to 2.85$\mu$m, average resolving power $R=700$) and order 2 (0.6 to 1.4$\mu$m, average resolving power $R=1400$). We therefore probe the full 0.6 to 2.85$\mu$m range at medium resolution \citep{albert_near_2023}. Our target is bright in the J band, which makes NIRISS/SOSS the only JWST mode that can be leveraged to probe potential near-infrared water bands. We only observe two groups up the ramp to avoid saturation, with a total of 756 integrations over each TSO. 

The first transit (Visit 1) was observed on UTC November 10, 2023, and covered 0.4 hours of baseline before the transit, 1.3 hours in transit, and the following 1.7 hours out of transit. The second transit observation (Visit 2) occurred on UTC November 16, 2023, and captured 1.4 hours of pretransit baseline, the 1.3-hour transit, as well as 0.7 hours of posttransit baseline. 

\subsection{\texttt{supreme-SPOON} reduction}\label{sec:supremespoon}

We reduce both SOSS visits using the \texttt{supreme-SPOON} pipeline\footnote{Now known as \texttt{exoTEDRF}.} \citep{Radica2024, feinstein_early_2023, radica_awesome_2023}, closely following the procedure in \citet{radica_muted_2024} and \citet{benneke_jwst_2024}. For each visit, we perform the standard \texttt{supreme-SPOON} Stage 1 and Stage 2 calibrations, including a group-level correction of column-correlated 1/$f$ noise, and a time-domain detection of cosmic rays specifically developed for ngroup=2 TSOs \citep{radica_muted_2024}. For the background, we perform a piecewise correction \citep[e.g.,][]{lim_atmospheric_2023, fournier-tondreau_near-infrared_2023, benneke_jwst_2024}, independently scaling the standard STScI SOSS background model on either side of the background ``step'' to match the flux level of the observations. Moreover, due to the presence of a dispersed contaminant located above the third diffraction order of the target spectrum in both observations, we use pixels $x \in [350, 600]$ and $y \in [230, 250]$ to estimate the pre-step background level instead of the default region. We extract the 2D stellar spectra using a simple aperture extraction with a full width of 40 pixels around the trace since the order self-contamination is predicted to be negligible \citep{darveau-bernier_atoca_2022, radica_applesoss_2022}.

\subsection{\texttt{NAMELESS} reduction}\label{sec:nameless}

We reduce the NIRISS/SOSS observations of GJ 9827\,d using the \texttt{NAMELESS} pipeline \citep{feinstein_early_2023,coulombe_broadband_2023} following the methods described in Coulombe et al. 2024 (submitted) and \citet{benneke_jwst_2024}. Our reduction starts from the uncalibrated raw data and goes through the super-bias subtraction, reference pixel correction, nonlinearity correction, ramp-fitting, and flat-fielding steps of the STScI \texttt{jwst} pipeline v1.12.5 \citep{bushouse_2023}. The jump detection step was automatically skipped by the pipeline as the integrations consist of only two groups. We then proceed with bad pixel correction where we flag and correct for pixels that show null, negative, or abnormally high counts at all integrations. The non-uniform background is subsequently subtracted by scaling independently the two regions of the STScI model background\footnote{Available at \url{https://jwst-docs.stsci.edu/}} which are separated by a sharp jump in flux around spectral pixel x$\sim$700, following the procedure presented in \citet{lim_atmospheric_2023}. We also correct for cosmic rays by computing the running median of the flux in each pixel, using a window size of 10 integrations, and clipping any 4$\sigma$ outliers. Finally, we correct for the 1/f noise by scaling all columns of the first and second spectral orders independently and considering only pixels less than 30 pixels away from the center of the trace, as described in detail in Coulombe et al. 2024 (submitted). The light-curves are then extracted from the first and second spectral orders using a simple box aperture with a width of 36 pixels.



\section{\texttt{ExoTEP} Light-curve fitting}\label{sec:lc_analysis}
We perform individual light-curve fits to each of the visits using the \texttt{ExoTEP} pipeline \citep{benneke_spitzer_2017,benneke_sub-neptune_2019}. We fit jointly the broadband light-curve fits of order 1 and order 2, to infer the system parameters, and then fix the planet's scaled semi-major axis $a/R_\star$ and its impact parameter $b$ to the best-fitting values from the joint white-light-curve fit when fitting wavelength-dependent light-curves. For NIRISS/SOSS, the \texttt{supreme-SPOON} extracted time-dependent spectra are used for the main analysis, and the resulting transmission spectrum is consistent with the one obtained with \texttt{NAMELESS}. 

\subsection{Broadband Light-curve fitting}

\begin{table*}
\caption{Stellar and planetary parameters adopted or refined in this work. The ephemeris is obtained by fitting a constant period and transit time to the observed \textit{Kepler}, \textit{HST}, \textit{Spitzer}, and \textit{NIRISS/SOSS} transit times together. The reference transit time is anchored to the epoch of the last JWST NIRISS/SOSS transit. The stellar luminosity is calculated from the literature values of the effective temperature and radius of GJ 9827. The planetary orbital separation, insolation, and equilibrium temperature values are derived from the scaled semi-major axis constraints obtained with NIRISS/SOSS, and using the literature values of the stellar radius and effective temperature. The planetary radius is obtained from the planet-to-star radius ratio fitted on the NIRISS/SOSS order 1 white light curves combined with the stellar radius constraint. }
\centering
\begin{tabular}{lcc}
\hline
\hline
Parameter      & Value             & Source         \\

\hline
\hline
GJ 9827 (star)&  & \\
\hline
Spectral type & K7V & \citet{dressing_characterizing_2019} \\
Effective temperature $T_\mathrm{eff}$ [K] & 4236 $\pm$ 12 & \citet{passegger_compact_2024} \\
Metallicity [Fe/H]& -0.29 $\pm$ 0.03 & \citet{passegger_compact_2024}\\
Age [Gyr]& 5.465 $\pm$ 4.058 &  \citet{passegger_compact_2024} \\
Stellar mass $M_\star$ [$M_\odot$]& 0.62 $\pm$ 0.04 & \citet{passegger_compact_2024} \\
Stellar radius $R_\star$ [$R_\odot$]& 0.58 $\pm$ 0.03 & \citet{passegger_compact_2024} \\
log$_{10}$ $\left(g_\star \left[cm/s^2\right]\right)$ [dex] & 4.70 $\pm$ 0.05 & \citet{passegger_compact_2024} \\
Stellar luminosity $L_\star$ [$L_\odot$]& $ 0.099 ^{+ 0.010 }_{- 0.009 }$ & Derived from lit. \\
Stellar rotation period [day]& 28.16$^{+3.38}_{-2.66}$ & \citet{passegger_compact_2024}\\

\hline
\hline
GJ 9827 d (planet) &  & \\
\hline
\textit{\textbf{Ephemeris}} &  & \\
Orbital period $P$ [days] & 6.201830 $\pm 0.000003  $ & This paper \\
Ref. transit time $T_0$ [BJD$_\mathrm{TDB}$] & $2460265.10196 \pm 0.00006$  & This paper \\
\textit{\textbf{Orbital parameters}} &  & \\
Scaled semi-major axis $a/R_*$   &   $19.739 ^{+ 0.619}_{ - 0.853}$ & This paper \\
Orbital separation $a_p$ [AU] & $ 0.053 \pm 0.003$& This paper   \\
Impact parameter $b$      &   $0.891 ^{+ 0.009}_{ - 0.010} $ & This paper \\ 
Inclination $i$ [deg] & $87.41 ^{+ 0.09 }_{- 0.11}$ & This paper \\
\textit{\textbf{Bulk planetary properties}} &  & \\
Planetary radius $R_p$ [$R_\oplus$]& $1.98 ^{+ 0.11 }_{- 0.10}$  & This paper   \\
Planetary mass $M_p$ [$M_\oplus$] & $3.02^{+0.58
}_{-0.57}$ & \citet{passegger_compact_2024}  \\
Insolation $S_\mathrm{p}$ [$S_\oplus$]& $ 34.9 ^{+ 2.8 }_{- 2.6 }$ & This paper  \\
Equilibrium temperature: &  & This paper     \\
~~~~--$T_\mathrm{eq,A_B = 0}$ [K] & $675 ^{+ 14 }_{- 12 }$& This paper     \\
~~~~--$T_\mathrm{eq,A_B = 0.3}$ [K] & $ 618 ^{+ 12 }_{- 9 }$ & This paper  \\
\hline
\label{table:pla_star_param}
\end{tabular}
\end{table*}


\begin{figure*}
    \centering
    \includegraphics[width=0.95\textwidth]{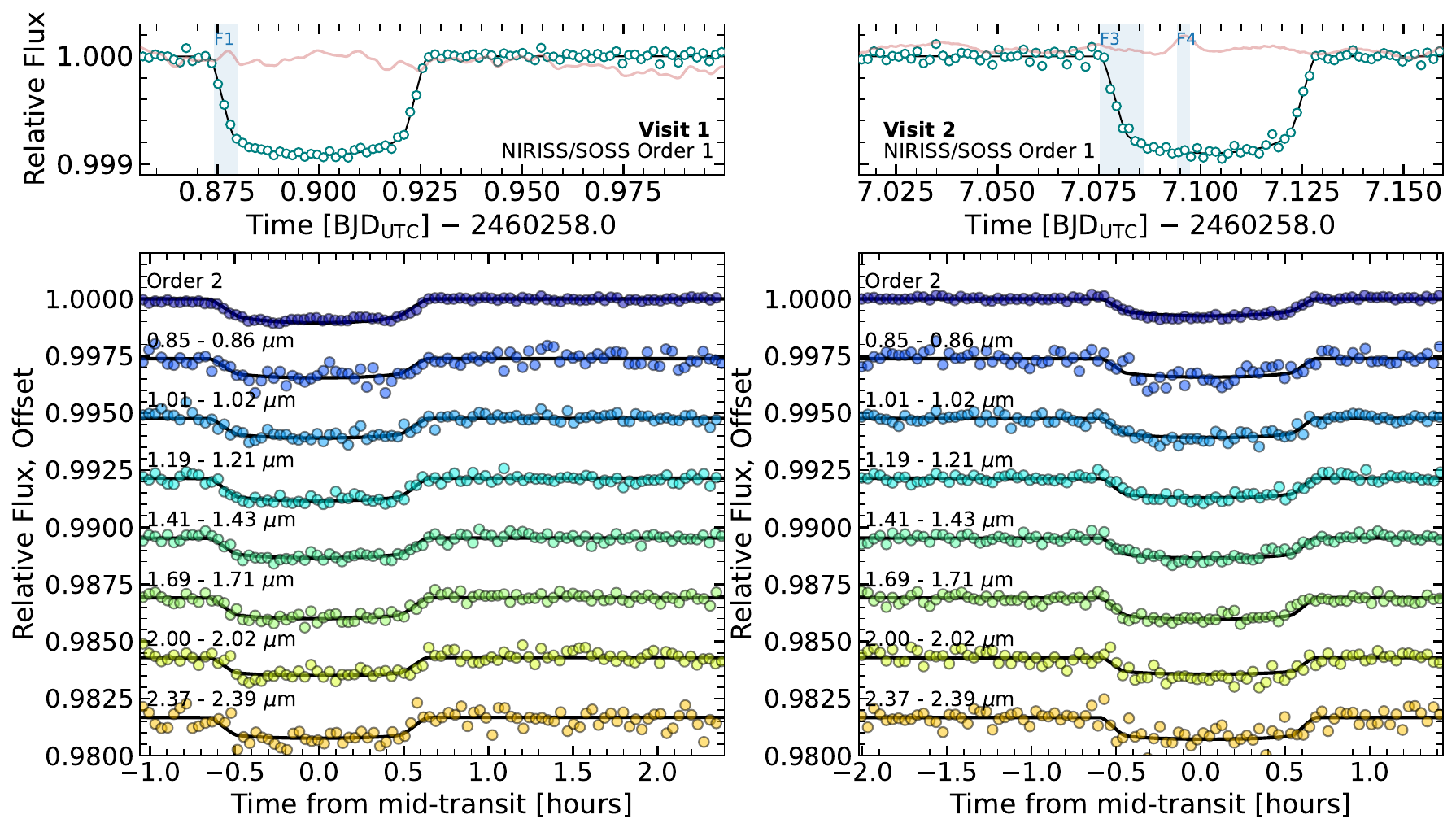}
    \vspace{5mm}
    \vspace{-5mm}\caption{\textit{Top panels:} Results from the white light curve fit of the supreme-SPOON reduction for Visits 1 (left) and 2 (right), and fitted systematics models. The order 1 systematics-corrected light-curves are shown (teal points), binned for visual purposes, along with the best-fit transit model (black) and the systematics model including the GP component (red, with the GP scaled for visualization purposes). The vertical blue shaded regions indicate the three candidate flares with the highest signal-to-noise spectra and following typical flare evolution profiles (see Section \ref{ssec:intransit_varia}).
    \textit{Bottom panels:} Spectroscopic light-curve fits to the two NIRISS/SOSS transits of GJ 9827 d from the \texttt{supreme-SPOON} reduction. The left panel corresponds to the first visit and the right panel to the second visit. We show the detrended broadband light-curve for order 2, and seven spectroscopic light-curves from order 1, from top to bottom. The best-fit astrophysical model is shown for each light-curve (black) and light-curves are offset relative to each other for clarity.
    }
    \label{fig:wlc_slc_fit}
\end{figure*}



For both TSOs, the NIRISS/SOSS white-light-curves for orders 1 and 2 exhibit the beating pattern that has been previously reported and is thought to correspond with the 2--4-minute cycling in the panel heating of the instrument electronics compartment \citep{mcelwain_james_2023}. These short-term variations are suppressed when detrending against the eigenvalues of the first two principal components extracted from the time series of detector images, which track changes in the trace morphology roughly corresponding to variations in the position (PC1) and width (PC2) of the trace as demonstrated in the eigenimages (see \citealp{coulombe_broadband_2023}). We found that even when detrending against the principal component analysis (PCA) components, residual correlated noise on longer timescales (10 to 30 min) is detected in the light-curves, which we tentatively attribute to stellar variability in the absence of corresponding variations in the recorded shifts of the spectrum or the trace width or position. We note, however, that since we are using a small number of groups up the ramp to accommodate the star's brightness, detector effects such as imperfect nonlinearity correction or outlier flagging that could affect time series observations of brighter targets might also have an impact on the extracted light-curves. We model this correlated noise source using the Gaussian Process (GP) kernel available in the \texttt{celerite} module that most closely reproduces a squared-exponential behavior (correlation between neighboring points over a characteristic timescale), the Matern 3/2 kernel.

Our final model combines the astrophysical (transit) component, as well as a systematics model. We use \texttt{ExoTiC-LD} \citep{david_grant_2022_7437681,husser_new_2013} to compute quadratic limb-darkening parameters, which we keep fixed (two parameters for order 1, two parameters for order 2) instead of fitting them, as the high impact parameter of GJ 9827 d poses a challenge to the joint fitting of the orbital and limb-darkening parameters. We use a custom SOSS order 2 throughput (O. Lim, priv. comm.) that covers all the extracted wavelengths, rather than the default \texttt{ExoTiC-LD} instrument throughput. Our systematics model has six free parameters: a linear trend with time, which is frequently observed in other JWST datasets (e.g., \citealt{coulombe_broadband_2023, feinstein_early_2023}), a linear function of the first two PCA eigenvalues approximately tracing vertical shifts and variations of the width of the trace, and residual correlated noise captured by the Matern 3/2 GP term, where we fit the correlation strength $\sigma_\mathrm{GP}$ and timescale $\tau_\mathrm{GP}$. We fit the white-light-curves of orders 1 and 2 together in order to extract the maximum information content on the orbital parameters of the planet that are fitted: the impact parameter $b$, scaled semi-major axis $a/R_\star$ and the transit time $T_c$. 

Beyond the previously mentioned systematics, visual inspection of the first NIRISS/SOSS visit reveals a jump in the light-curve around $BJD_\mathrm{TDB}=2460258.91$ (in the transit) with a sudden increase followed by a decrease in flux, and a sudden decrease in flux close to $BJD_\mathrm{TDB}=2460258.955$ (in the post-transit baseline). 
Besides, the light-curve of the second NIRISS/SOSS visit displays a short in-transit bump, which we investigate in more detail and attribute to a stellar flare (see Section \ref{ssec:intransit_varia}). To model these visit-specific effects independently, we perform separate white-light-curve fits for the first and second visits.

While the small in-transit bump observed in visit 2 does not bias the determination of the system parameters from the light-curve as the egress remains unaffected by the flare, the feature in the transit for visit 1 is much broader and hinders independent determination of $b$ and $a/R_\star$, as it cannot be efficiently masked while obtaining sufficient phase coverage of the transit. Therefore, we first fit the cleaner light-curve of the second visit and extract from their posterior distributions the covariance matrix for $b$ and $a/R_\star$, which we use as a correlated Gaussian prior on these two parameters when fitting the light-curve of the first visit (Figure \ref{fig:wlc_slc_fit}).

For the second visit, we model the in-transit bump by adding a narrow Gaussian bump to the systematics model, which has three more parameters: the center of the Gaussian $t_\mathrm{sc}$ [BJD$_\mathrm{TDB}$], its width $\sigma_\mathrm{sc}$ and its amplitude $A_\mathrm{sc}$ (see Roy et al. 2024, in prep).

Before performing the final fit to the white light curves of visit 1, we obtained a Gaussian prior on its transit time $T_\mathrm{c,1}$ from a simple fit to the portions of the light-curve in both orders that appeared least affected by stellar or instrumental nuisance signals (excluding BJD $\in [2460258.85, 2460258.87]$ and BJD$\geq 2460258.955$). For this simple fit, the systematics model did not include a GP and only accounted for a linear detrending with time and with the eigenvalues from the first two principal components. Finally, we ran a full fit to the entire light-curve, introducing again the Matern 3/2 GP component, and using a covariance matrix to encode our prior on the values of $T_\mathrm{c,1}$, $b$, and $a/R_\star$ (Figure \ref{fig:wlc_slc_fit}).

We find consistent planetary parameters across both visits (Table \ref{table:wlc_param}) and in agreement with literature values (Table \ref{table:pla_star_param}; Section \ref{sec:comp_lit}). We also obtain an updated ephemeris for GJ 9827 d (see Table \ref{table:pla_star_param}; Section \ref{sec:comp_lit}), and do not detect significant transit-timing variations (TTVs) using the two additional transit time measurements (see Figure \ref{fig:TTVs}; Section \ref{sec:comp_lit}).

\subsection{Spectroscopic light-curve fitting}

We obtain the transmission spectrum of GJ 9827 d for each visit by fitting the spectroscopic light-curves using \texttt{ExoTEP} \citep{benneke_spitzer_2017}, and fixing the orbital parameters ($b$, $a/R_\star$), and the mid-transit time $T_c$ to the best-fit parameters from the white light curve. We bin the extracted 2D spectra to obtain three versions of the transmission spectrum. The first version of the spectrum, at a resolving power of 100, is used in the retrieval analyses as it provides a good compromise between computationally expensive retrievals on a higher-resolution spectrum and loss of spectral information for heavier wavelength binning (Welbanks et al., in prep). We also produce two higher-resolution spectra where we only extracted the order 1 spectrum around the metastable He I triplet to search for any He absorption feature. For these He search analyses, we use one version of the spectrum obtained by fitting the transit depth in every detector column and one where we bin together two neighboring columns. We verify that all three versions of the spectrum agree in the spectral region around the metastable He triplet where the high-resolution spectra were obtained (Figure \ref{fig:compare_three_spectra}).

Similarly to the white light curve analysis, we fit jointly the astrophysical model described by the planet-to-star radius ratio $R_p/R_\star$, and a systematics model including parameters that capture instrumental and astrophysical noise, to each spectroscopic light-curve. Because of the high impact parameter of the planet (Table \ref{table:pla_star_param}), we keep the limb-darkening parameters fixed to the 1D quadratic parameters computed in each bin by ExoTIC-LD using the stellar parameters of GJ 9827. We fit to each light-curve a scatter term $s$, the two parameters of a slope with time $c + v t$, and slopes with the eigenvalues of the first two principal components, $v_\mathrm{PC1}$ and $v_\mathrm{PC2}$. For visit 2, we additionally fit the amplitude and width of a Gaussian describing the bump, with the feature center time fixed to the best-fit value from the white light curve fit. 

We try fitting GP parameters to each spectroscopic light-curve but find that they are unconstrained in most spectroscopic bins and lead to overestimated transit depth uncertainties. Rather, we elect to use the recorded GP model hyperparameters and the corresponding mean GP model of the residuals from the best fit to the white light curve. For the first visit, we fix the Matern 3/2 GP lengthscale to the best-fit white light curve values and model the residuals in each spectroscopic bin by only fitting for the correlation strength parameter $\sigma$ over a narrow range of values (factor of 0 to 10) around the best-fitting value from the white light curve fit. For the second visit, where the structure of the residuals in each spectroscopic bin match more closely those of the white light curve, we use the mean GP model computed from the GP parameters of the white light curve best-fit systematics model and scale it by fitting a multiplicative factor in each bin. We check that the residuals from the fitted spectroscopic light-curves (Figure \ref{fig:wlc_slc_fit}) follow the distribution expected from white noise using Allan plots.

The visit 2 spectrum is robust to the choice of GP model for the spectroscopic light-curve fit as the scaled white light curve model or the scaled best-fit GP model parameters to the white light curve, and we report the spectrum obtained using the less-flexible approach \citep[e.g.,][]{radica_muted_2024}. For both visits, we confirm that the spectrum we obtain with this approach is consistent within less than 1$\sigma$ with the spectrum obtained from the more conservative approach of excluding the regions of the light-curve that are directly within the in-transit bump features.





We repeat the same white- and spectroscopic light-curve fitting procedure for the \texttt{supreme-SPOON} reduction of both visits, which produces a spectrum consistent with \texttt{NAMELESS} at better than $1\sigma$ (Figure \ref{fig:compare_reductions}). We note, however, that there are slight differences between the two reductions in how well the extracted principal components correlate to the white light curve residuals. The correlation is stronger for the \texttt{NAMELESS} reductions, which leads to a lower residual scatter. 
The overall slope in the white light curve baseline is also different between the two reductions, but it does not affect the resulting spectrum as the small differences (with a mean difference of 11 ppm overall) are absorbed by the slope and GP parameters. 

Our spectrum is also in excellent agreement with the spectrum obtained with \textit{HST} (Figure \ref{fig:pipeline_comparison}) from combining 10 transits of GJ 9827 d \citep{roy_water_2023}, with a small vertical shift which we attribute to the fact that the orbital parameters $b$ and $a/R_\star$ were fixed to literature values \citep{niraula_three_2017} due to partial ingress/egress coverage when fitting the \textit{HST} data. 

We calculate the final spectrum by averaging the spectra from the two visits and use the spectrum from the \texttt{supreme-SPOON} reduction for the atmospheric analysis (Table \ref{tab:trans_sp}).

\subsection{Interpretation of the observed stellar variability features}\label{ssec:interp_varia}

We study the wavelength dependence and time evolution of the residual features in the light-curves that are captured by the GP in our fit (Section \ref{ssec:intransit_varia}). Although the time resolution offered by NIRISS/SOSS does not allow us to probe the detailed time evolution of each feature (Figure \ref{fig:flare_timeseries}), we prefer flares to stellar spot crossings because of the ``fast-rise, exponential decay'' (FRED) evolution observed in the time series of flare effective temperatures that we derive from the observations for three of the candidate flare events, labeled F1, F3, and F4 (Figures \ref{fig:wlc_slc_fit} and \ref{fig:flare_timeseries}). We do not detect H$\alpha$ variability, which was found to be a telltale flare diagnostic for the much cooler star TRAPPIST-1 \citep{lim_atmospheric_2023,howard_characterizing_2023}, and attribute this to the difference in brightness and spectral energy distribution between the M7.5V star TRAPPIST-1 and the K7V star GJ 9827, where flares may be characterized by continuum enhancements rather than H$\alpha$ variations (see Figure \ref{fig:example_flare_spectra}; discussion in Section \ref{ssec:intransit_varia}). The short-wavelength coverage of NIRISS/SOSS enables us to extract, for each of the candidate flare events, the flare energy in the \textit{TESS} bandpass, and to leverage relations between the TESS flare energy and the near-UV and far-UV (NUV, FUV) energies to estimate the NUV and FUV outputs corresponding to each candidate flare (Table \ref{tab:flares_data}).



\section{Atmosphere and stellar contamination modeling} \label{sec:atm_modeling}

We use SCARLET and POSEIDON to perform 1D retrievals of the atmospheric makeup of GJ 9827 d from the NIRISS/SOSS and WFC3 transmission spectra. We also consider the potential impact of unocculted stellar heterogeneities on the inferred atmospheric properties by fitting the spectrum assuming that it can be entirely explained by stellar contamination, or accounting for the contributions of both the planetary atmosphere and any potential stellar surface heterogeneities on the observed features.


\subsection{Atmosphere modeling with SCARLET}\label{sec:scarlet}

We perform both free chemistry and chemically consistent retrievals over the spectrum of GJ 9827 d using the SCARLET atmospheric retrieval framework \citep{benneke_atmospheric_2012,benneke_how_2013,benneke_strict_2015,benneke_sub-neptune_2019, benneke_water_2019, pelletier_where_2021, piaulet_evidence_2023}. 

The version of SCARLET used in this study leverages several upgrades over previously published work, including the implementation of the joint fitting of a stellar contamination contribution to the spectrum using the \texttt{MSG} module \citep{townsend_msg_2023} alongside the atmosphere parameters, and the introduction of partial aerosol coverage using
a cloud fraction \citep{welbanks2019a}. In chemically consistent retrievals, we also run cases where we allow the CH$_4$ and/or NH$_3$ abundance to be fitted freely independent of the chemical equilibrium prediction, to test the robustness of our results and potential disequilibrium depletions and avoid unphysically low C/O ratios if e.g. methane is not detected. 

We explored a range of assumptions to explain the spectrum of GJ 9827 d, from free retrievals with well-mixed abundances of the fitted species to chemical equilibrium retrievals that follow the predicted molecular abundances in thermochemical equilibrium for the fitted isothermal temperature, atmospheric metallicity, and C/O ratio, with or without a freely fitted CH$_4$ or NH$_3$ abundance to capture potential disequilibrium depletion.

For free retrievals, the forward model of SCARLET performs the radiative transfer and iterative hydrostatic equilibrium calculation for the atmospheric composition and temperature structure prescribed by the sampled parameters, while for chemically consistent retrievals the molecular abundances are set not only by the sampled metallicity and C/O ratio, but also by the local pressure and temperature conditions in each layer of the atmosphere. Our free retrieval includes H$_2$, He, N$_2$, HCN, H$_2$O, CO, CO$_2$, CH$_4$, NH$_3$, H$_2$S, and for absorbing species, we use the HELIOS-K computed cross sections of H$_2$O \citep{ExoMol_H2O}, 
 CO \citep{Hargreaves2019}, CO$_2$ \citep{ExoMol_CO2},  CH$_4$ \citep{HargreavesEtal2020apjsHitempCH4}, HCN \citep{Harris_HCN_2006}, NH$_3$ \citep{ExoMol_NH3}, H$_2$S \citep{ExoMol_H2S}. 

For each combination of model parameters, the radius at the reference pressure of 10 mbar is fitted to find the value that provides the best match between the model spectrum and the observations, with the hydrostatic equilibrium and radiative transfer steps iterated over until the optimal radius value is found. 

The chemical composition of the planet is parameterized either by the fitted abundances (free retrieval) or by the log atmospheric metallicity, C/O ratio, and CH$_4$ abundances (chemically consistent retrievals). We fit for the temperature $T_p$ of the photosphere which is sampled by our observations. For the aerosols, we use either a gray cloud and hazes or a more complex Mie cloud parameterization \citep{benneke_sub-neptune_2019}. In the first scenario, clouds and hazes are parameterized respectively by the log of the cloud-top pressure (approximately $\tau=1$ level), and $c_\mathrm{haze}$, the slope enhancement parameter relative to Rayleigh scattering, caused by small particles. 
In the Mie cloud scenario, we fit for the scale height of the cloud relative to the atmosphere, H$_\mathrm{rel}$, for the density of cloud particles $n$ and for the mean particle size $\mu$. 
We allow for patchy clouds with the $f_\mathrm{cloud}$ parameters, where clouds are patchy if $f_\mathrm{cloud}<1$. The parameter space is sampled using multi-ellipsoid nested sampling as implemented in the \texttt{nestle} module \citep{skilling_nested_2004, skilling_nested_2006}. Models are computed at a resolving power of 15,625 and then convolved to the resolution of the observed spectrum for the likelihood calculation. We additionally fit for a potential offset between the Order 1 and Order 2 transmission spectrum and between the Order 1 NIRISS/SOSS and \textit{HST}/WFC3 spectra.

For the temperature profile, we fit the temperature of the photosphere of the atmosphere at the terminator, which is representative of the region probed by our transmission observations. In free retrievals, we fit for the abundances of the molecular species that have spectral features within the NIRISS/SOSS bandpass, with wide log-uniform priors on the volume mixing ratios (lower bound at $10^{-10}$): H$_2$O, CH$_4$, and NH$_3$, as well as CO, CO$_2$,  H$_2$S, and HCN which have much weaker expected opacity over the NIRISS/SOSS wavelength range (few absorption bands and/or low expected volume mixing ratios). We ran retrievals where the background, or filler gas is H$_2$/He (enforcing a Jupiter-like He/H$_2$ = 0.157), H$_2$O, or N$_2$.

\subsection{Atmosphere modeling with POSEIDON}\label{sec:poseidon}

We use the combined transmission spectrum of GJ~9827~d using both JWST/NIRISS SOSS data and HST/WFC3 data from \citet{roy_water_2023} to infer properties of the atmosphere.
We do this by also conducting atmospheric retrievals with the open-source retrieval code POSEIDON \citep{macdonald_hd_209458b_2017, macdonald_poseidon_2023}, which utilizes the nested sampling algorithm PyMultiNest \citep{feroz_multinest_2009, buchner_x-ray_2014} to investigate the potential atmospheric models, from which we use 2000 live points for each of our models.
The descriptions of the radiative transfer technique, forward atmospheric modeling, and the opacity database that POSEIDON employs can be found in \citet{macdonald_trident_2022}.
From our data, we produce spectra from 0.6 to 2.9 $\mu$m at R = 20000, which is convolved with a Gaussian kernel to the native resolution of NIRISS/SOSS (R $\approx$ 650) and HST/WFC3 (R $\approx$ 150), respectively, and then multiplied by the instrument sensitivity function and subsequently binned down to the wavelength spacing of our data.

We consider a wide range of atmospheric models, including a flat line model, pure H$_2$O atmospheres, multigas atmospheres, and with or without stellar contamination (see Section \ref{sec:joint_retrieval_poseidon}). Our multigas models include the following gases: H$_2$, He, N$_2$, HCN, H$_2$O, CO, CO$_2$, CH$_4$, NH$_3$, and H$_2$S. 
We note that our POSEIDON multigas models also use a permutation-invariant centered-log-ratio (CLR) prior \citep{benneke_atmospheric_2012} for the volume mixing ratio, meaning any molecule is theoretically equally likely to be the dominant gas.
We fit for a reference radius, R$_\mathrm{p,ref}$, at a reference pressure of 10 bars, assuming an isothermal atmospheric temperature.
All of our non-flat models include both clouds and hazes, where we assume an optically-thick gray opacity, and all layers deeper than P$_\mathrm{surf}$ are set to infinite opacity. We also use a two-parameter power-law prescription for hazes \citep{macdonald_hd_209458b_2017} in all of our multigas atmospheric models.
The priors we use in our retrievals are as follows: R$_\mathrm{p,ref}$ (R$_\oplus$) $=$ $\mathcal{U}$ (0.60, 1.15), T (K) $=$ $\mathcal{U}$ (100, 1000), log a$_\mathrm{haze}$ $=$ $\mathcal{U}$ (-4, 8), $\gamma_\mathrm{haze}$ $=$ $\mathcal{U}$ (-20, 2), log P$_\mathrm{cloud}$ $=$ $\mathcal{U}$ (-7, 2), log X $=$ $\mathcal{CLR}$ (-12, 0).






\subsection{Stellar-contamination-only retrievals with \texttt{stctm}}\label{sec:stctm_methods}

We use the open-source \texttt{stctm} module \citep{piaulet_stctm_2024} previously applied to TRAPPIST-1 b \citep{lim_atmospheric_2023}, TRAPPIST-1 g (Benneke et al., submitted), and  GJ 9827 d \citep{roy_water_2023} to model the case where the planet's atmospheric signature would have no wavelength dependence. This corresponds to a scenario where the planetary atmosphere contribution to GJ 9827 d's transmission spectrum would be featureless, e.g., from high-altitude opaque uniform clouds, or in a bare rock scenario which is unlikely given the planet's low bulk density.

In order to check whether stellar surface heterogeneities can completely explain the water features in GJ 9827 d's spectrum, we use \texttt{stctm} to perform stellar contamination-only retrievals of the transmission spectrum assuming that (1) only spots contribute, and (2) that both spots and faculae play a role (parameterization described in Appendix Section \ref{sec:stellar_ctm}). 

In the spots-only case, we fit for $\Delta T_\mathrm{spot}<0$ (see Appendix Section \ref{sec:stellar_ctm}), for the spot covering fraction $f_\mathrm{spot}$, and the planet's wavelength-independent transit depth $D$. In the spots+faculae case, we fit for two additional parameters: $\Delta T_\mathrm{fac}>0$ and $f_\mathrm{fac}$ which describe the temperature and covering fraction of faculae. 
To obtain the specific intensity $I_{\lambda}$ of the spots, faculae, and photosphere, we use the \texttt{MSG} module \citep{townsend_msg_2023} which interpolates across the PHOENIX grid of stellar atmosphere models \citep{husser_new_2013}. To speed up each likelihood evaluation, we pre-compute a grid of stellar models at a resolving power of 10,000 with \texttt{MSG}, which match the star's surface gravity $\log~g$, and span the effective temperatures $T_\mathrm{eff}$ in small intervals of 20\,K. In the retrieval, we use the stellar model within the grid corresponding to the closest $T_\mathrm{eff}$ to the sampled value of $T_\mathrm{spot}$, $T_\mathrm{fac}$ or $T_\mathrm{phot}$.

The retrievals leverage the \texttt{emcee} implementation of Markov-Chain Monte Carlo sampling to explore the parameter space \citep{foreman-mackey_emcee_2013}. We use 20 times as many walkers as there are fitted parameters, and run the chains for 5,000 steps, discarding 60\% as burn-in prior to obtaining the final posterior distributions on the heterogeneity parameters. We use uniform priors on the heterogeneity fractions and $\Delta T_\mathrm{het}$, with $|\Delta T_\mathrm{het}|>100$ as otherwise the heterogeneity spectra are so similar to the photosphere spectrum that any heterogeneity fractions can be invoked. We jointly fit for the photosphere temperature $T_\mathrm{phot}$ to propagate its uncertainty, with a Gaussian prior informed by the literature constraint (Table \ref{table:pla_star_param}, \citealt{passegger_compact_2024}). The faculae fraction and temperature being largely unconstrained in the second retrieval configuration, without improving the quality of the fit to the data, we only present results for the spot-only fit.

\subsection{Joint stellar contamination and planetary atmosphere retrievals with SCARLET}\label{sec:joint_retrieval_scarlet}

We perform atmospheric retrievals with SCARLET using a conservative treatment of the potential transit light source (TLS) effect on the spectrum where we marginalize over any contributions consistent with the observations. We use the SCARLET implementation of joint atmosphere and stellar contamination retrievals that was developed for the study of HAT-P-18\,b \citep{fournier-tondreau_near-infrared_2023}, and update the stellar modeling query component to use the \texttt{MSG} module instead of the standard \texttt{pysynphot} implementation. 

For these joint retrievals, we therefore simultaneously sample the parameters of the planetary atmosphere and of the stellar surface, as parameterized in the \texttt{stctm} implementation. The stellar contamination factor $\epsilon_{\lambda, \, \rm{het}}$ is multiplied by the predicted transmission spectrum given the sampled atmosphere parameters prior to the likelihood evaluation. We test cases where we consider no stellar contamination, spots only (with a single population having a shared $T_{\rm{spot}}$), and both spots and faculae with an associated temperature for each heterogeneity population.

\subsection{Joint stellar contamination and planetary atmosphere retrievals with POSEIDON}\label{sec:joint_retrieval_poseidon}


We further use POSEIDON to determine the degree to which the TLS effect may be able to explain GJ~9827~d’s transmission spectrum in the absence of an atmosphere.
POSEIDON produces model spectra by multiplying a bare-rock transmission term, (R$_p$/R$_*$)$^2$, by the wavelength-dependent stellar contamination factor from two distinct stellar heterogeneities (see Equation \ref{eq:stellar_contam_factor}).
Using the \texttt{PyMSG} package, POSEIDON calculates the spectra of the active regions by interpolating across the PHOENIX grid of stellar atmosphere models \citep{husser_new_2013}. 
We add five additional free parameters to each of our TLS models, the priors of which are as follows: f$_{fac}$ $=$ $\mathcal{U}$ (0.0, 0.5), f$_{spot}$ $=$ $\mathcal{U}$ (0.0, 0.5), T$_{fac}$ (K) $=$ $\mathcal{U}$ (T$_*$-36, 1.2T$_*$), T$_{spot}$ (K) $=$ $\mathcal{U}$ (2300, T$_*$+36), T$_{phot}$ (K) $=$ $\mathcal{U}$ (T$_*$, 12), where T$_*$ $=$ 4236 $\pm$ 12.
The surface gravity of the stellar regions is fixed to log g $=$ 4.719 $\pm$ 0.02 \citep{passegger_compact_2024}.








\section{Atmospheric escape modeling}\label{sec:escape_modeling}

We examine hydrodynamic escape in more detail, using the CHAIN (Cloudy e Hydro Ancora INsieme) 1D hydrodynamic model introduced in \citet{kubyshkina2024A&A...684A..26K} (see details of the modeling setup in Appendix Section \ref{sec:chain_description}). 

We compute mass loss rates, as well as temperature, pressure, density, and ionization profiles with height for several composition cases. We take as the reference case a H$_2$/He-dominated atmosphere with a metallicity matching that of the star and also run models for water-enriched H$_2$/He atmospheres (with varying mass fractions of water mixed with the hydrogen and helium, see \citealt{egger_unveiling_2024} for more detail), or H$_2$/He atmospheres with increased metallicity relative to solar (following the same methodology as \citealt{Linssen2024high_metallicities}). The water- and metal-enriched simulations allow us to account for the effect of enrichment in heavier species on the stability of the atmosphere against escape.

\section{Planetary structure modeling}\label{sec:structure_modeling}

We use \texttt{smint} \citep{piaulet_wasp-107bs_2021,piaulet_evidence_2023}, an open-source Bayesian estimator of planetary compositions, to constrain the properties of GJ 9827 d in the scenario where water is the only volatile present on top of a rocky core with an unconstrained ratio of silicates to iron. The model used accounts for the various phase transitions of water depending on the temperature and pressure conditions \citep{aguichine_mass-radius_2021} which, for GJ 9827 d, corresponds to water in the gas, then in the supercritical phase from low to high pressures due to its irradiation level. 

We use a Gaussian prior on the irradiation temperature of GJ 9827 d ($T_\mathrm{irr}=674 \pm 29$\,K), calculated from Monte Carlo sampling of the stellar effective temperature, radius, and the planetary orbital distance (Table \ref{table:pla_star_param}) and on the planet mass \citep{passegger_compact_2024}. We uniformly sample the fraction of iron in the iron core + silicate mantle interior of the planet,  $f'_\mathrm{core}$ ($f'_\mathrm{core}=0$ in the absence of an iron core, $f'_\mathrm{core}=325$ for an Earth-like interior composition) and use a uniform prior on the water mass fraction $f_\mathrm{H_2O}$. The parameter space is sampled using \texttt{emcee} \citep{foreman-mackey_emcee_2013}. We run the chains for 2,000 steps and discard the first 60\% as burn-in to obtain the final posterior distributions.


\section{Results} \label{sec:atm_results}

\begin{figure*}
    \centering
    \includegraphics[width=0.9\textwidth]{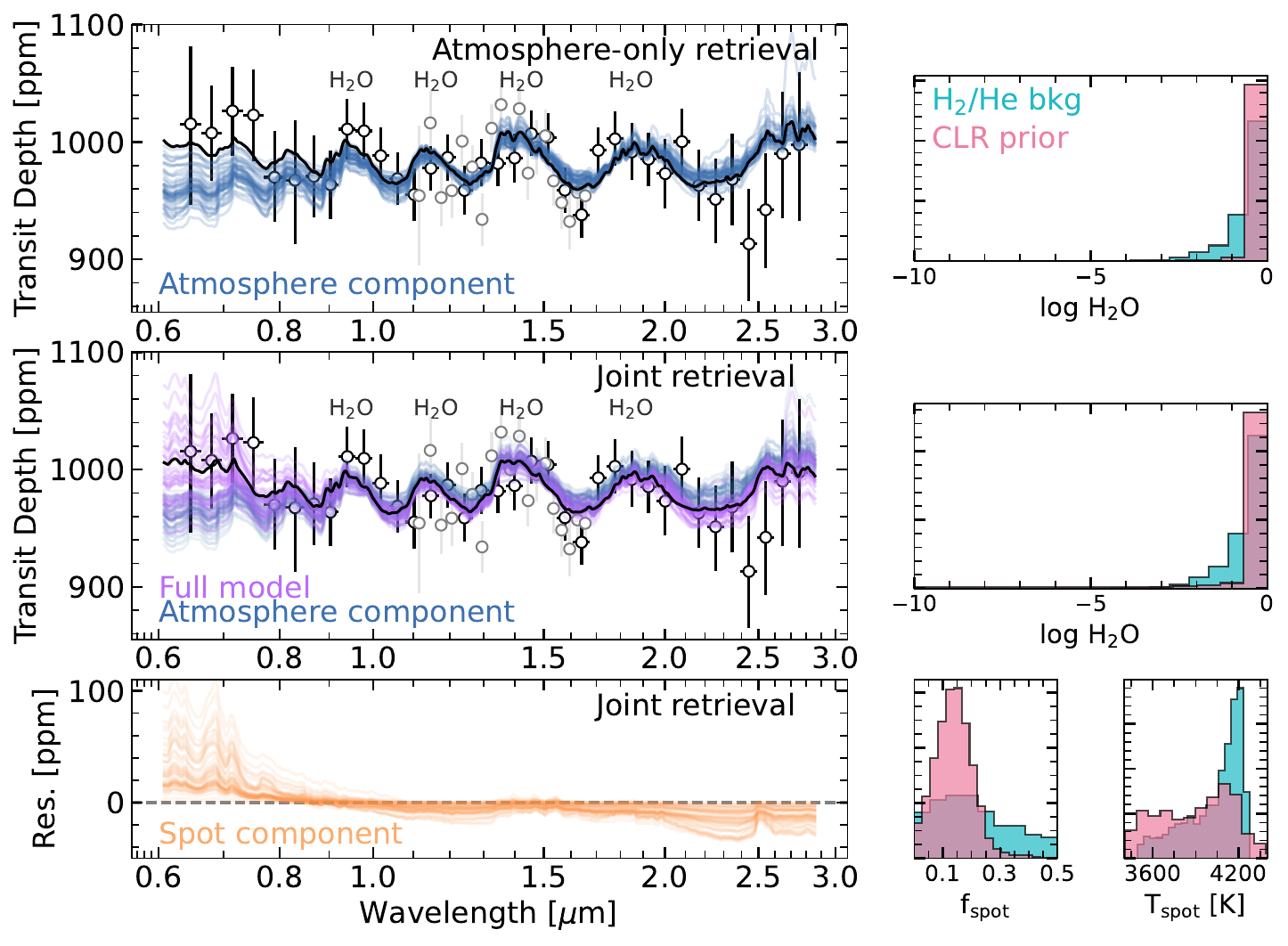}
    \vspace{5mm}
    \vspace{-5mm}\caption{Results from the free retrieval performed on the JWST/NIRISS SOSS + HST/WFC3 spectrum of GJ 9827 d.
    \textit{Left panels:} The semi-transparent lines are sample spectra from the posterior distributions of the SCARLET free retrievals when only the atmosphere (top panel) or both the atmosphere and stellar contamination component (bottom two panels) are accounted for. The black line is the best-fit model, and the black (gray) points are the measured NIRISS/SOSS (HST/WFC3) transit depths. The retrievals were performed using the R=100 version of the JWST/NIRISS SOSS spectrum, and a binned version is shown for visual comparison purposes. Even when stellar contamination is accounted for, the planetary atmosphere component of the model captures the H$_2$O absorption features, and the constraints on the H$_2$O abundance are not affected. The model including the contribution from spots only introduces an additional slope and does not affect the retrieval results. 
    \textit{Right panels:} Posterior distributions obtained on H$_2$O in both retrieval setups, and on the spot covering fraction and temperature in the joint retrieval. For the POSEIDON retrievals (CLR prior) we only show in this figure the samples where the spot is at most 800 K cooler than the photosphere, for clarity (see full posteriors in Figure \ref{fig:1d_distri_retrieval}). Our inference of a water-rich, steam atmosphere is robust to the marginalization over stellar contamination contributions, and to the choice of priors (H$_2$/He as the background gas in blue, CLR prior in pink). }
    \label{fig:retrieval_spectrum_fit}
\end{figure*}

\begin{figure*}
    \centering
    \includegraphics[width=0.9\textwidth]{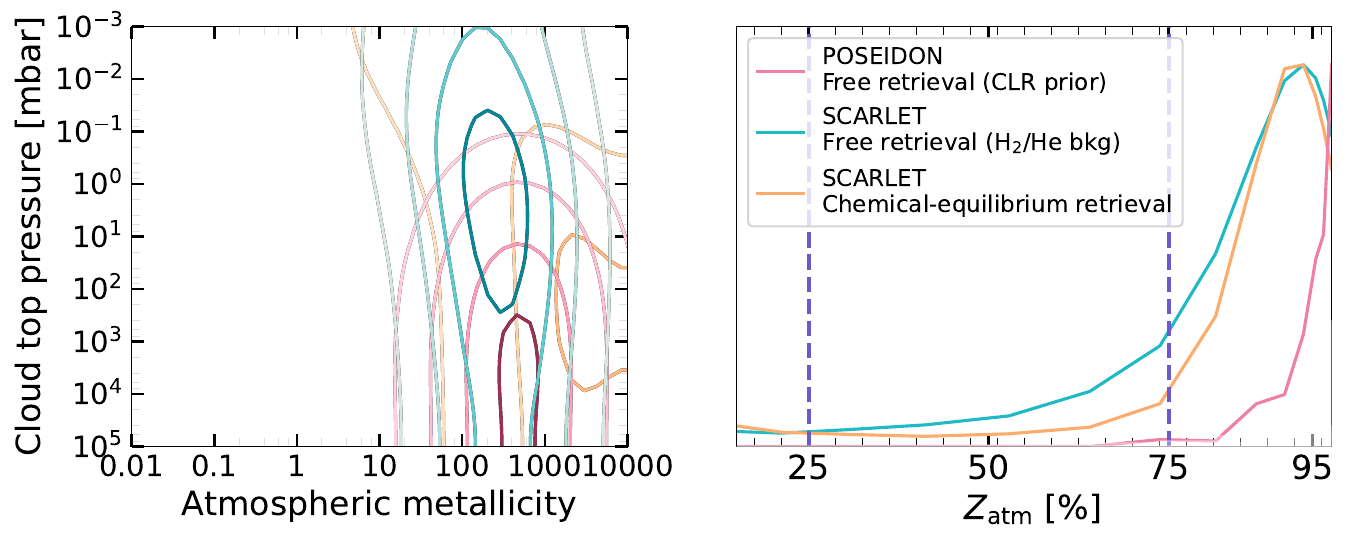}
    \vspace{5mm}
    \vspace{-5mm}\caption{Constraints on the atmospheric composition from the POSEIDON retrieval with the agnostic (CLR) prior on the background gas (pink), the SCARLET retrieval with H$_2$/He assumed to be the background gas (blue), and SCARLET chemically-consistent retrieval (orange). \textit{Left panel:} Contours representing the 0.5, 1, 1.5, and 2$\sigma$ levels in the joint posterior probability distribution on the atmospheric metallicity and the gray cloud top pressure, for all three retrievals. \textit{Right panel:} Kernel density estimations of  the atmospheric metal mass fraction $Z_\mathrm{atm}$ obtained from all posterior samples, for each retrieval setup.}
    \label{fig:met_pcloud_zatm}
\end{figure*}

We perform retrievals to jointly interpret the HST/WFC3 \citep{roy_water_2023} and our NIRISS/SOSS transmission spectrum of GJ 9827 d (Figure \ref{fig:retrieval_spectrum_fit}), which cannot be explained by stellar contamination alone (Figure \ref{fig:stctm_retrieved_spectrum}). Atmospheric retrievals favor a high mean molecular weight, water-rich atmosphere (Figures \ref{fig:1d_distri_retrieval},\ref{fig:met_pcloud_zatm}).
We also search for a signature of escaping metastable helium by leveraging the high spectral resolution in NIRISS/SOSS order 1 but do not detect any significant signal. Our hydrodynamic simulations predict atmosphere mass loss rates much higher than our measured upper limit. Overall, the data are best explained by a   ``steam world'' atmosphere with a high water abundance. 

\subsection{Constraints on the atmospheric composition}\label{sec:atm_composition}

We fit the JWST/NIRISS SOSS + HST/WFC3 transmission spectra of GJ 9827d jointly, as the large number of transits observed with HST (10) yielded small enough error bars to be comparable in precision with the NIRISS/SOSS spectrum (Figure \ref{fig:pipeline_comparison}).
We find that the absorption features in the transmission spectrum of GJ 9827 d can be robustly attributed to water vapor in its atmosphere, as stellar spots alone cannot reproduce the amplitude of the observed features (Figure \ref{fig:stctm_retrieved_spectrum}), in agreement with the HST results \citep{roy_water_2023}. We do not detect any other molecular species. Most of the posterior probability mass of free retrievals regardless of the assumed background gas (Figure \ref{fig:1d_distri_retrieval}) and of the chemically-consistent retrievals (Figure \ref{fig:met_pcloud_zatm}) lies in a metal-enriched regime with dozens of percent of the atmosphere mass contained in metals (Figure \ref{fig:met_pcloud_zatm}). 

We find that a high mean-molecular weight water-rich atmosphere can reproduce the spectrum even without any cloud opacity (Figure \ref{fig:met_pcloud_zatm}). Free retrievals performed with SCARLET and POSEIDON favor high water abundances, with tentative upper limits on the CH$_4$ and NH$_3$ abundances driven by the nondetection of the 2.3\,$\mu$m CH$_4$ features (other features overlap with water bands), and weak constraints on the 1.5 and 2.0\,$\mu$m ammonia features (Figure \ref{fig:retrieval_spectrum_fit}). CO and CO$_2$ abundances are not significantly constrained by our observations (Figures \ref{fig:vmr_constraints_allmols}, \ref{fig:vmr_constraints_allmols_clr}). Retrievals with CLR priors specifically favor H$_2$O rather than H$_2$ as the dominant species in the atmosphere. 

With H$_2$O as the only O-bearing molecular species with detectable bands in the NIRISS/SOSS bandpass and no detection of any C-bearing species, our constraints, especially on the C/H ratio, remain wide (Table \ref{tab:mol_constr}), while the measured O/H ratio is precise even when marginalizing over undetected CO or CO$_2$. Potential SO$_2$ in the atmosphere (with no significant absorption in the NIRISS/SOSS bandpass) may alter our O/H estimate. The most agnostic retrieval setup run (CLR free retrieval) measures a mean atmospheric molecular weight that closely matches that of water ($18.01^{+0.20}_{-0.61}$) and estimates that about four times as much mass is contained in O as in H.

Our constraints on the atmospheric composition are robust to the choice of filler gas in the SCARLET retrievals: H$_2$/He (top panels in Figure \ref{fig:1d_distri_retrieval}), H$_2$O, N$_2$ (not shown), or an equal prior on any molecule being the filler species (centered-log-ratios retrievals performed with POSEIDON, see bottom panels in Figure \ref{fig:1d_distri_retrieval}). The interpretation is consistent across the assumptions made, from chemical equilibrium to free retrievals, with small variations related to the prior on how molecular abundances vary with height (Figures \ref{fig:1d_distri_retrieval} and \ref{fig:met_pcloud_zatm}). We note that even when using log uniform priors for the volume mixing ratio (H$_2$/He assumed as the background ``filler'' gas), we find similar strong evidence of a H$_2$O-dominated atmosphere, in full agreement with the SCARLET results.

The Mie cloud parameterization of particle size distributions \citep{benneke_sub-neptune_2019} was found not to be warranted by the quality of our data, from a model comparison standpoint, and all the retrievals presented model aerosols as a gray cloud and a potential haze slope in all the results discussed. Our data only marginally ($\sim 1.9\sigma$) favor CH$_4$ depletion compared to chemical equilibrium (Table \ref{table:model comparison}). 

\begin{table*}
\caption{\label{tab:mol_constr} Atmospheric retrieval results for GJ 9827 d. We quote results from the free retrievals including one population of stellar heterogeneity (spots) with POSEIDON (CLR prior case, for the samples where $T_\mathrm{spot}$ is no more than 800\,K cooler than $T_\mathrm{phot,star}$) and SCARLET (H$_2$/He as filler gas case). The uncertainties represent the 68\% confidence intervals for all two-sided constraints and 95\% (2$\sigma$) upper or lower limits otherwise.}
\centering
\begin{tabular}{lccc}
\hline
\hline
Parameter & Agnostic& H$_2$/He \\
 & background gas & background gas \\
\hline
\multicolumn{3}{l}{\textbf{Fitted atmosphere parameters}} &  \\
\textit{Detection} &  &  \\
         \logX{H2O} & $>-0.50$ ($>31.62$\%) & $>-1.86$ ($>1.38$\%)\\ 
\hline
\textit{Upper limits} &  &  \\

         \logX{CH4} & $<-3.60$ & $<-3.71$\\ 
         \logX{NH3} & $<-3.55$ & $<-3.68$\\ 
         \logX{H2S} & $<-2.67$ & $<-2.68$\\ 
         \logX{CO2} & $<-1.69$ & $<-1.83$\\ 
         \logX{CO}  & $<-0.90$ & $<-0.82$\\ 
         \logX{HCN} & $<-2.67$ & $<-2.46$\\     
\hline
\hline
\multicolumn{3}{l}{\textbf{Fitted stellar contamination parameters}} &  \\
T$_\mathrm{phot, star}$ [K] & $4235.45^{+11.56}_{-11.19}$ & $4305.36_{-10.2}^{+10.7}$\\ 

T$_\mathrm{spot}$ [K]\footnote{For the SCARLET retrieval, derived from the samples on $T_\mathrm{phot}$ and $\Delta T_\mathrm{spot}$} & $3942.56^{+230.3}_{-345.28}$ & $4102.5_{-315.8}^{+108.0}$\\ 

f$_\mathrm{spot}$ [K] & $0.13^{+0.13}_{-0.08}$ & $ 0.23 _{-0.15}^{+ 0.35}$\\

\hline
\hline
\multicolumn{3}{l}{\textbf{Fitted instrumental offset parameters}} & \\
NIRISS/SOSS order 2 offset & N/A & $-4.4_{-22.6}^{+20.89}$\\
HST/WFC3 offset & $-15.88^{+6.42}_{-6.53}$ & $15.7_{-4.6}^{+5.0}$\\

\hline
\hline

\multicolumn{3}{l}{\textbf{Derived quantities}} & \\

          MMW [amu] & $18.02^{+0.20}_{-0.61}$ & $9.84_{-4.64}^{+4.51}$\\
$\logXratio{CH4}{H2O}$ & $<3.52$ & $<-3.11$\\ 
$\logXratio{NH3}{H2O}$ & $<3.44$ & $<-3.03$\\ 
  O/H by number & $0.499_{-0.007}^{+ 0.008}$ & $ 0.25_{ -0.17}^{+0.15} $\\ 
    O/H by mass & $4.344 \pm 0.069$ & $2.18_{-1.48}^{+1.28}$\\ 
  log$_{10}$ C/H by number & $-3.8 \pm 2.0$ & $-4 \pm 2$\\ 
   log$_{10}$ C/H by mass & $-2.8 \pm 2.0$ & $-2.9 \pm 2.0$\\ 
log$_{10}$ (C+O)/H by mass & $0.638_{-0.004}^{+0.019}$ & $0.35_{-0.50}^{+0.2}$\\ 
            $Z_\mathrm{atm}$\ [\%] & $>97.2$ & $>25.2$\\ 
\hline
\hline
\end{tabular}
\end{table*}


\subsection{Impact of stellar contamination and instrument offsets} \label{sec:impact_stctm_offsets}

We investigate the impact of accounting for potential stellar surface heterogeneities on the results from our retrievals. We show that for both free and chemically consistent retrievals, the addition of the contamination signal from spots (cooler regions on the stellar surface) and faculae (hotter plages) is not significantly favored from a Bayesian model comparison standpoint (Table \ref{table:model comparison}). Further, we demonstrate that the addition of spots, or both spots and faculae, does not alter our inference of atmospheric properties (Figure \ref{fig:1d_distri_retrieval}). This finding is in line with previous \textit{HST}/WFC3 results \citep{roy_water_2023} which rule out stellar contamination as the source for the 1.4$\mu$m spectral band that was detected with \textit{HST}. 

In particular, contrary to M dwarfs where spots can often easily reproduce planetary atmosphere-like water bands, the K7 dwarf GJ 9827 would require unrealistically large covering fractions of spots over $\sim$1000\,K cooler than the quiet photosphere to produce the observed 1.4$\mu$m water absorption features (see modeling performed in \citealt{roy_water_2023}, and Figure \ref{fig:stctm_retrieved_spectrum}). Modeling performed with POSEIDON for TLS-only models agrees with the \texttt{stctm} models and cannot reproduce the observed features.

When stellar contamination is considered in addition to a planetary atmosphere, testing for spots, spots+faculae, and without any stellar contamination for Bayesian comparison, we only find weak Bayesian evidence for the spot-only model, while we find no evidence of stellar faculae (Table \ref{table:model comparison}). We try independently fitting each NIRISS/SOSS visit with or without spots or faculae to be robust to the presence of different stellar contamination contributions in each visit's spectrum, but the addition of stellar heterogeneity is disfavored from a model comparison standpoint in each case.

We also fit for a vertical offset between the NIRISS/SOSS order 1 and order 2 transmission spectra, and we find that the retrieved offset is very small compared to the error bars on the individual transit depths. For example, the H$_2$/He background free retrieval with stellar spots finds an offset of $-4^{+21}_{-20}$ ppm. However, the addition of an offset (retrieved to be $15.7^{+5.0}_{-4.6}$ for the atmosphere + spots case) is essential to jointly fit the JWST and HST data due to the difference between the assumed (HST) and fitted (JWST) orbital parameters of the planet (see Table \ref{tab:mol_constr}).

\subsection{Search for escaping metastable helium}

\begin{figure}
    \centering
    \includegraphics[width=0.45\textwidth]{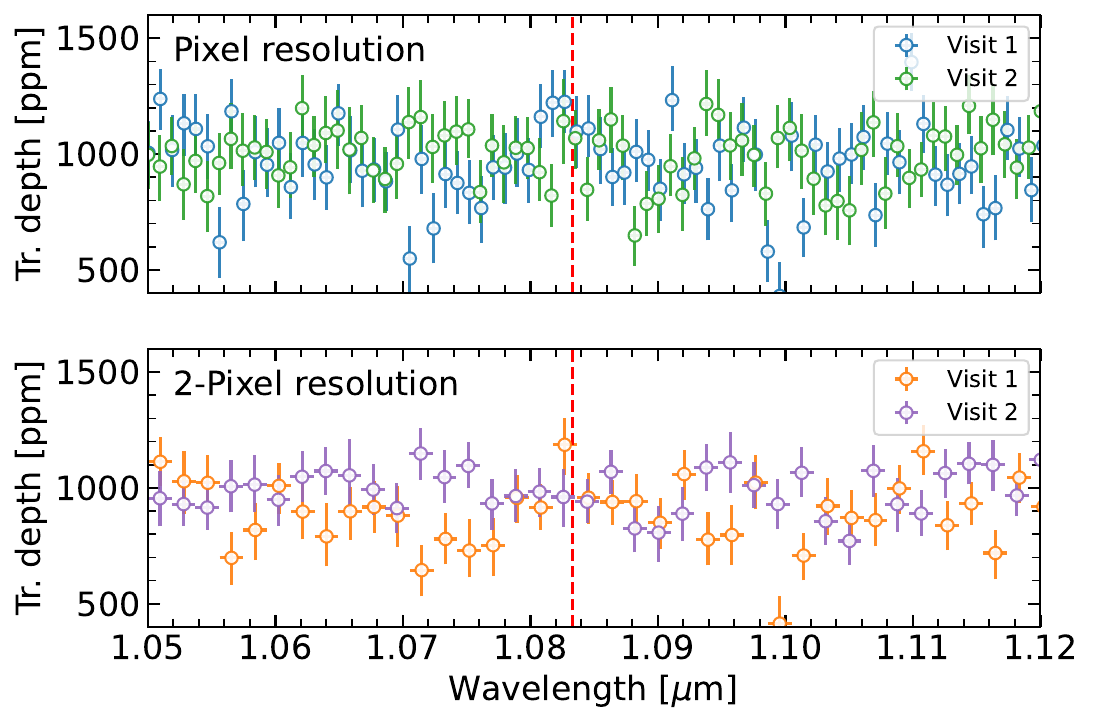}
    \vspace{5mm}
    \vspace{-5mm}\caption{Search for metastable He absorption around 10,833 \AA\ in both NIRISS/SOSS transits of GJ 9827 d. The top and bottom panels respectively show the spectrum fitted after extraction at pixel resolution, and 2-pixel resolution, for each of the two visits (\texttt{supreme-SPOON} reduction). We do not detect significant absorption from the He I metastable triplet, in line with more sensitive ground-based high-resolution observations.}
    \label{fig:hetriplet_search}
\end{figure}

The first order of NIRISS/SOSS covers the metastable He I triplet, which could show up in the transmission spectrum as a result of an extended escaping H/He atmosphere on GJ 9827 d. We sought to leverage the resolving power of NIRISS/SOSS to search for a He I signature. We produced two new spectral extractions, one at full resolution, and one at 2-pixel resolution, in the vicinity of the He I triplet (Figure \ref{fig:hetriplet_search}). We fit the full resolution spectrum a Gaussian with a fixed width (0.75\,\AA) convolved at the resolution of NIRISS/SOSS (700), and at the expected location of the He I triplet absorption (10833.33\,\AA), to constrain the sensitivity of our observations to a potential signal.

We do not report any detection of the helium triplet in the NIRISS/SOSS full-resolution spectrum of GJ 9827 d. We obtain a constraint on the amplitude of the helium absorption signal in our data which corresponds to an upper limit on the amplitude of a helium absorption signal of  $0.56 \pm 0.23$ \% at high spectral resolution (80,000).
The lack of metastable helium absorption on GJ 9827 d is surprising, as planets in similar irradiation conditions to GJ 9827 d around K-type stars are more likely to have helium particles in their metastable state leading to the detection of extended thermosphere and escaping material \citep{oklopcic_helium_2019}. Still, our upper limit is consistent with more constraining nondetections of escaping metastable He and neutral H obtained from the ground with Keck/NIRPSEC, CARMENES or the InfraRed Doppler (﻿IRD) instrument on the Subaru telescope \citep{kasper_non-detection_2020,carleo_multiwavelength_2021,krishnamurthy_absence_2023}.

\subsection{Hydrodynamic escape predictions}

For the hydrogen-helium dominated atmosphere simulation with a stellar metallicity, our model predicts an atmospheric mass loss rate of $\sim2\times10^{10}$~${\rm g\,s^{-1}}$, and the escape is dominated by photoionization heating from atomic hydrogen. 
The ion fraction in the H$_2$/He dominated atmosphere remains relatively low ($f_{\rm ion}$ is only $\sim 0.2$ at the Roche lobe), suggesting that hydrogen escape should be detectable in Ly$\alpha$ line \citep[see e.g.,][]{kubyshkina2022MNRAS.510.2111K}.

For water-enriched atmosphere cases compatible with the range of atmospheric water enrichments indicated by the transmission spectrum of GJ 9827 d (Figure \ref{fig:met_pcloud_zatm}), we obtain a reduction in the atmospheric mass loss rates by a factor of $\sim2$ (for a water mass fraction of $70$\%) and up to an order of magnitude (factor of $\sim8$ for a water mass fraction of $95$\%). The reduction is similar in the high-metallicity atmosphere cases, with slightly lower mass loss rates compared to water-rich atmospheres for the same atmospheric mean molecular weight, due to heating from metal ions. While the increase in mean molecular weight is the main driver for the decrease in mass loss rate, oxygen line cooling (for the water-rich cases) and cooling via the metal lines (for the high-metallicity cases) can contribute up to 90\% of the total radiative cooling in metal-enriched atmospheres.

We also find that even though a large fraction of the escaping atmospheric material still consists of atomic hydrogen in water-rich atmospheres, the increasing mean molecular weight leads to more compact atmospheres with 
narrower ionization fronts (the ion fraction increases faster with altitude). Therefore, neutral hydrogen particles are limited to low altitudes, and the total ion fraction in the outflow increases by up to a factor of a few hundred at the sonic point. We note, however, that our models that use the Cloudy line lists are not conclusive as to the impact of increased metallicity on the detectability of the metastable He feature beyond the fact that helium itself would be less abundant in atmospheres composed largely of water, or of metals more generally.

Finally, our models predict that for water-rich atmospheres (but much less so for high-metallicity atmospheres), a large fraction of the escaping material is represented by atomic oxygen supplied by the photodissociation of water molecules above $\mu$bar pressure level. At high altitude, most of the oxygen is expected to be in ion form, with O$^{+}$ and O$^{2+}$ making up as much as 70\% of the outflow by mass at the sonic point (up to $\sim$3 orders of magnitude more than in H$_2$/He dominated atmospheres; see \citealp{egger_unveiling_2024}). It remains unclear, however, whether these oxygen species would escape along with the hydrogen (see Section \ref{sec:chain_description}).


\subsection{Structure modeling results}

If the entire volatile content of GJ 9827 d was contained within a pure water vapor/supercritical envelope (Figure \ref{fig:smint_results}), we find that its water mass fraction would be $32 \pm 10$\% when marginalizing over the full range of plausible iron fractions in the planetary interior. This constrains to approximately 60--80\% the total mass of the planet locked in its rock/iron core, assuming the mantle has completely solidified. 

Our water mass fraction is consistent with the 5--30\% reported by \citet{aguichine_mass-radius_2021} for GJ 9827 d. The differences between the two studies can be traced back to the fact that \citet{aguichine_mass-radius_2021} used a previous determination of the planetary parameters with a higher mass for GJ 9827 d \citep{rice_masses_2019}, and did not marginalize over the uncertainty on the planet's irradiation temperature. 

\section{Discussion}\label{sec:discussion}

\subsection{GJ 9827 d as a mixed-envelope H$_2$O-rich ``steam world''}\label{sec:mixed_water_world}

The most likely interpretation for the high-metallicity atmosphere with abundant water revealed by all the fits we performed is a ``steam world'' mixed atmosphere scenario, with dozens of percent of the atmosphere comprised of high mean-molecular weight species (Figure \ref{fig:met_pcloud_zatm}). For the warm equilibrium temperature of GJ 9827 d, we expect its temperature-pressure profile to cross directly from the gas to the supercritical phase of water, without allowing for any water condensation in the atmosphere or at the surface. Since hydrogen and water are highly miscible in the gas and supercritical phases \citep{soubiran_miscibility_2015, pierrehumbert_runaway_2022,innes_runaway_2023}, HMMW volatiles can be expected to be well-mixed throughout the envelope, and the measured upper-atmosphere O/H ratio can be expected to be representative of the deep envelope, with the ratio of H$_2$O/CO/CO$_2$ dictated by the local thermodynamical conditions \citep{benneke_jwst_2024}. The measured O/H ratio only provides a lower limit on the oxygen content accumulated by the planet during its formation, as more refractories can be locked in the deep interior \citep{schlichting_chemical_2021,dorn_hidden_2021,luo_majority_2024}.

\subsection{nondetection of atmospheric escape in the context of a metal-rich atmosphere}


Beyond the atmospheric composition constraints, our hydrodynamic simulations reveal that a metal-rich atmosphere is favored for GJ 9827 d because of the instability of a hydrogen-dominated atmosphere to atmospheric escape and the nondetection of escaping neutral hydrogen. 

While our hydrodynamic simulations for a hydrogen-rich atmosphere scenario predict that the escape would be detectable in Ly $\alpha$, observations obtained only upper limits \citep{kasper_non-detection_2020,carleo_multiwavelength_2021,krishnamurthy_absence_2023}. Furthermore, even at the present-day modeled mass loss rate, the H$_2$-He atmosphere would be removed in less than 1~Gyr, while the atmospheric mass loss rates at the beginning of evolution are expected to be 10--100 times higher \citep{kubyshkina2024A&A...684A..26K}.
Therefore, a hydrogen-helium atmosphere is inconsistent with the age of the GJ~9827 system \citep[which is $\geq$1.4~Gyr][]{passegger_compact_2024}.

However, the mass loss rate is significantly reduced in metal-enriched and steam atmosphere cases.  We find that if the atmosphere metallicity is greater than 500$\times$ solar, or for water mass fractions of $\gtrsim90$\%, the mass loss is reduced enough to shield the planet from photoevaporative stripping.  Moreover, high water mass fractions also lead to an increase by two orders of magnitude in the fraction of hydrogen atoms escaping in the ionized form rather than in the neutral form. This can complicate the detection of hydrogen escape with methods like absorption in the Ly$\alpha$ line, which targets neutral hydrogen. 

The escape and atmospheric evolution argument therefore further supports the ``steam world'' scenario with a water-rich, high metallicity atmosphere favored by the atmospheric retrieval. In this scenario, and since GJ 9827 d is not a rocky planet from bulk density considerations, the nondetection of an escaping atmosphere has two plausible explanations: the escape rate is too low to be detected (for GJ~9827~d, the expected reduction compared to the H-He dominated atmospheres is up to an order of magnitude), or H and He escape in different states that are not probed by the neutral H (Lyman $\alpha$) and metastable He transitions where this search has been conducted to date. 

\subsection{Constraints on the interior composition of a water-rich GJ 9827 d}

We infer a bulk water mass fraction of $32 \pm 10$\% for GJ 9827 d in the case of a water steam atmosphere, which is larger than that of the more massive, larger sub-Neptune TOI-270 d \citep{benneke_jwst_2024}. GJ 9827 d would not only have a more metal-rich atmosphere, but it may also have formed from the accretion of a lower relative amount of rocky vs. icy material, as the rock/iron mass fraction inferred for TOI-270 d is $\sim 90$\%.

We note that our approach could underestimate the total water budget of GJ 9827 d, as significant amounts of water can remain locked in the core or mantle of the planet during mantle crystallization through volatile partitioning\citep{salvador_convective_2023}, or could still be dissolved in the magma ocean if at least some of the mantle still is in a molten state \citep{dorn_hidden_2021}. Another important consideration is the possibility that other volatiles such as CO$_2$ could be abundant in the planetary atmosphere, making it more compact than this simple model predicts and altering the inferred mass locked in the core and mantle. 

\subsection{Volatile enrichment mechanisms}


There are three main pathways to a volatile-rich, and specifically a water-rich atmosphere: accretion of ice-rich material during or after planet formation, interactions between a magma ocean and the planetary atmosphere, or gradual increase in the atmospheric metallicity from selective loss of lighter species. If GJ 9827 d formed beyond the ice line, it could have accreted large amounts of water ice in solids \citep{leger_new_2004,luger_habitable_2015,alibert_formation_2017,kite_habitability_2018,venturini_nature_2020,bitsch_dry_2021}. As the planet migrates inward, the photosphere will then become warm enough that a fraction of the accreted water will partition in the vapor form in the atmosphere where it can be probed observationally. Alternatively, magma ocean geochemical interactions with a hydrogen-rich atmosphere can provide an endogenous source for tens of percent of water in the planetary atmosphere, or atmospheric metal mass fractions of $\sim 30-70$\% \citep{schlichting_chemical_2021,kite_water_2021,rogers_fleeting_2024}.
A third pathway to enhance the metal content of an initially hydrogen-dominated atmosphere which may have been accreted in-situ is the preferential loss of hydrogen over heavier metal species via mass fractionation \citep{hu_helium_2015,chen_evolutionary_2016,malsky_coupled_2020}. This process requires a stronger gravitational well for GJ 9827 d than for cooler planets which are more traditionally believed to be in the ``sweet spot'' to balance the effect of metals being lost alongside hydrogen. We find that on the $\sim 6.2$d orbit of GJ 9827 d, its mass of 3.02 $M_\oplus$ is greater than the $\gtrsim2.5 M_\oplus$ required to undergo mass fractionation in their atmospheres if the dominant mode of atmospheric escape is core-powered mass loss. However, the higher escape fluxes expected from EUV-driven photoevaporation would drive the loss of the bulk of the metal species in the escaping flow \citep{cherubim_strong_2024}.  If the temperature is high enough, this can lead to the drag of heavier metals and, therefore, fractionation. Recent research by Louca et al. (in prep.) demonstrated that planets originating as warm super-Neptunes can undergo significant mass loss, shedding both hydrogen and heavier metals. This process results in high-metallicity atmospheres for mature sub-Neptune-like planets. Typically, most of the hydrogen-dominated envelope is lost within the first billion years post-formation, leaving these planets with a considerably lower escape rate in their later stages (see e.g., \citealt{Lopez2012}; \citealt{Kubyshkina2020}).



\subsection{A potentially oxidizing atmosphere}

\begin{figure*}
	\centering
	{
		\includegraphics[width=0.8\textwidth]{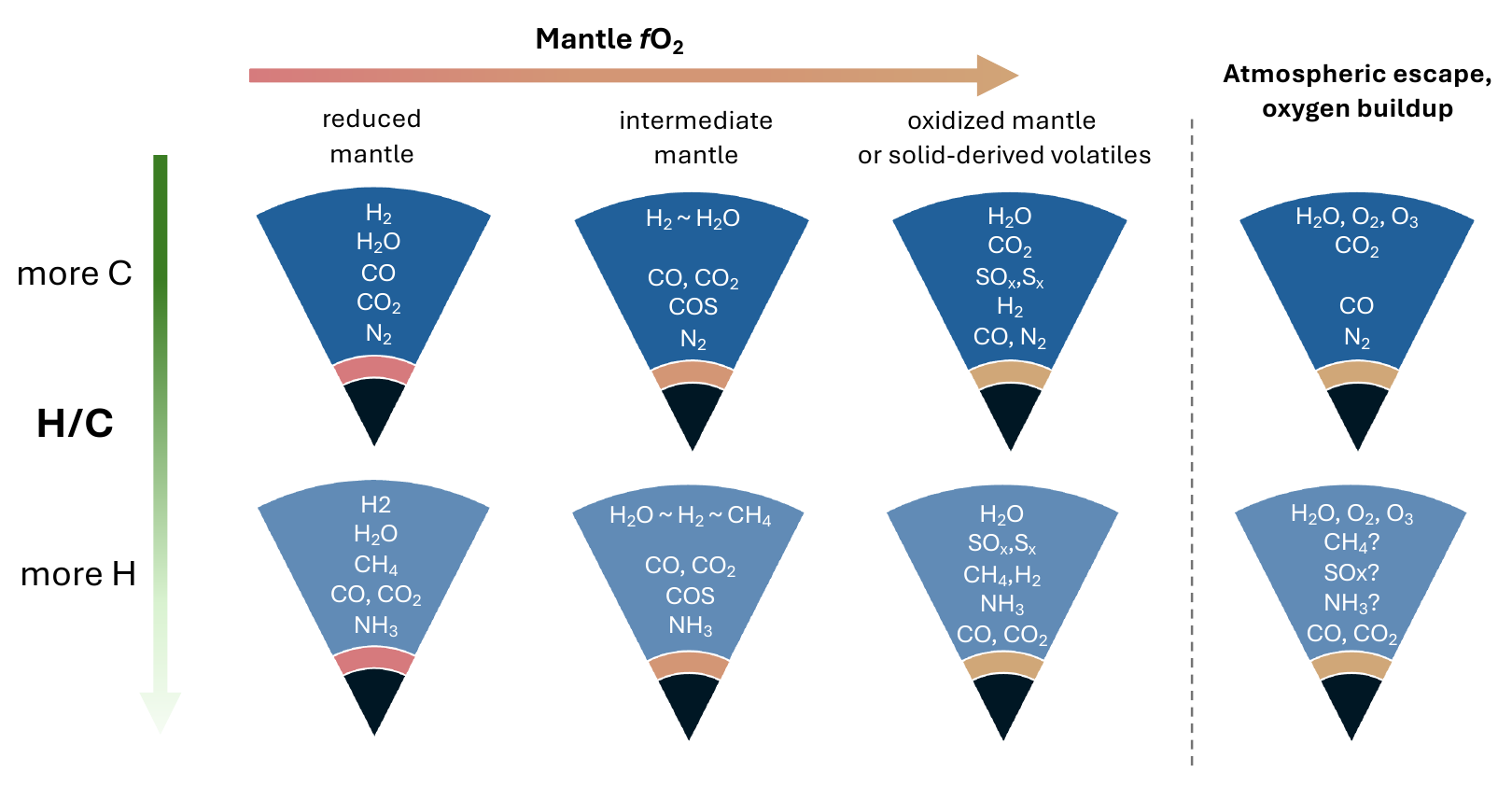}
	}
	\caption{Qualitative description of the major oxygen-, carbon-, nitrogen- and sulfur-bearing species expected in metal-rich atmospheres as a function of atmospheric origin and mantle composition (white labels inside of each mantle composition case). Water-dominated atmospheres can result from magma ocean-atmosphere interactions under oxidized mantle conditions \citep{schlichting_chemical_2022,kite_atmosphere_2020} or from the planet-forming or accreted solids that coagulated beyond the water ice line. Depending on the redox state of the mantle and the H/C ratio of the mantle \citep{liggins_growth_2021}, CH$_4$ or CO/CO$_2$ can be the dominant carbon-bearing species, and NH$_3$ or N$_2$ can be the dominant nitrogen-bearing species. Further O$_2$ buildup can occur over time as H is gradually lost to space \citep{krissansen-totton_oxygen_2021,harman_snowball_2022}. }
	\label{fig:comp_HMMWV}
\end{figure*}

The high water mixing ratio suggested by our \textit{HST}/WFC3 + \textit{JWST}/NIRISS SOSS joint retrieval raises the possibility of a more oxidizing bulk atmosphere composition, which could result from an oxidized mantle composition, the accretion of solid-derived volatiles, or even gradual oxygen buildup over time (Figure \ref{fig:comp_HMMWV}). 

Oxygen-rich atmospheres have been predicted for temperate water worlds due to the suppression of oxygen sinks by overburden pressure \citep{krissansen-totton_oxygen_2021}. They have also been proposed as a possible evolutionary outcome for warm sub-Neptunes interior to the runaway greenhouse limit due to extensive water dissociation and H escape \citep{harman_snowball_2022}. Nebular hydrogen in contact with molten silicates will reduce FeO to produce metallic Fe and H$_2$O \citep{kite_water_2021}; the resulting atmosphere may possess a broad range of H$_2$:H$_2$O inventories depending on nebular endowment, initial mantle FeO, extent of equilibration, and removal of metallic Fe in the core \citep{schlichting_chemical_2022,kite_atmosphere_2020}. In contrast, the accretion of predominantly icy material onto silicates will result in a more oxidized (higher H$_2$O:H$_2$ ratio) bulk atmosphere composition \citep{kite_atmosphere_2020}. For these planets, a plausible amount of H escape could overwhelm mantle oxygen sinks and flip atmospheric composition from net reducing (H$_2$O-H$_2$ dominated) to oxidizing (H$_2$O-O$_2$ dominated) \citep{harman_snowball_2022}. The absence of reduced species like CH$_4$ and NH$_3$ (Figure \ref{fig:1d_distri_retrieval}) and an aerosol-free atmosphere due to the photochemical oxidation of haze-forming species could be consistent with such an atmosphere. The lack of sensitivity to CO/CO$_2$ over the NIRISS/SOSS wavelength range, however, precludes us from a precise estimation of the total oxygen content in the atmosphere.

\begin{figure}
    \centering
    \includegraphics[width=0.48\textwidth]{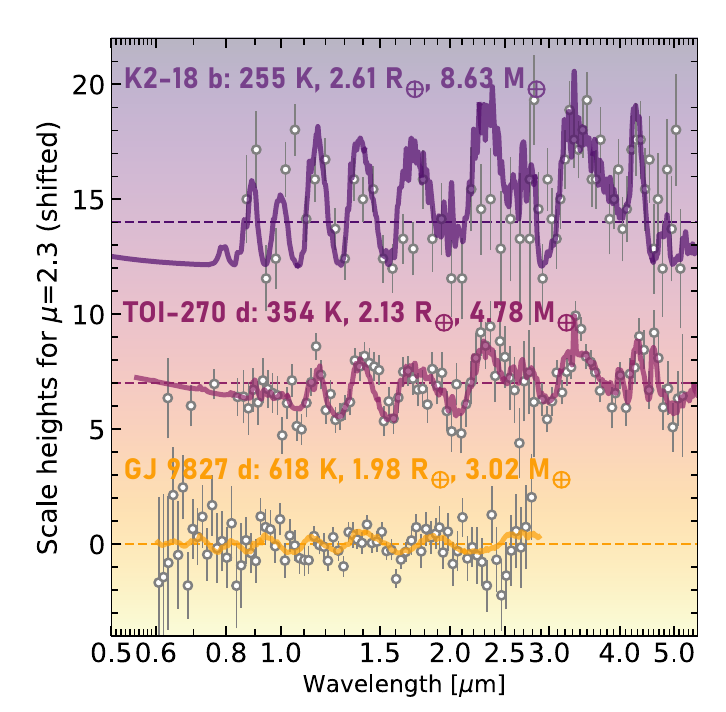}
    \vspace{5mm}
    \vspace{-5mm}\caption{Transmission spectra of K2-18 b \citep{madhusudhan_carbon-bearing_2023}, TOI-270 d \citep{benneke_jwst_2024}, and GJ 9827 d, in units of scale heights for a H$_2$/He-dominated atmosphere, offset. For clarity, the spectra of TOI-270 d and GJ 9827 d are binned to a resolving power of 50 (gray points). The best-fit models from the retrievals run on the spectra of GJ 9827 d and TOI-270 d are shown on top of the observations (colored lines) and for K2-18 b we show a representative SCARLET model for the retrieved abundances quoted in \citet{madhusudhan_carbon-bearing_2023}. The background color and line colors encode the variation in equilibrium temperature between the three sub-Neptunes, but they also have different sizes and densities (see annotated temperatures for a Bond albedo of 0.3, masses, and radii). }
    \label{fig:comparative_exoplanetology}
\end{figure}

\subsection{Evidence for compositional diversity in sub-Neptune atmospheres} \label{sec:compositional_transitions_subNep}

Our characterization of the atmosphere of GJ 9827 d underscores the diversity in the atmospheres of small sub-Neptunes where initial C/N/O/S volatile endowments, (potentially selective) atmospheric escape, and interior-atmosphere interactions over the planet's thermal evolution can all significantly alter the resulting atmosphere. We illustrate this diversity by visually comparing the transmission spectra of TOI-270 d and GJ 9827 d with the larger K2-18 b (Figure \ref{fig:comparative_exoplanetology}). While K2-18 b was found to have a hydrogen-dominated, low mean molecular weight atmosphere, with a typical ``gas dwarf'' composition \citep{madhusudhan_carbon-bearing_2023}, the transmission spectra of TOI-270 d and GJ 9827 d both reveal much more metal-enriched atmospheres \citep{benneke_jwst_2024} as reflected in their smaller vertical extent on Figure \ref{fig:comparative_exoplanetology} where they are scaled relative to the number of planetary scale heights spanned by the spectral features (calculated for a typical H$_2$/He dominated atmosphere). 

The trend of decreasing, then increasing spectral feature strength for planets with warm-temperate equilibrium temperatures, with a ``minimum'' found near $\sim 600$\,K \citep{brande_clouds_2023} is also reflected in the gradual decrease in feature strength from the colder K2-18 b to the warmer TOI-270 d and GJ 9827 d. However, while initially attributed to the effect of vertically-extended clouds, or small-particle hazes as a function of equilibrium temperature \citep{brande_clouds_2023}, clouds were \textit{not} significantly detected in the atmospheres of TOI-270 d or GJ 9827 d, with the variety of spectral feature strengths seemingly arising from differences in atmospheric mean molecular weight, rather than cloudiness trends. This motivates further exploration of the diverse atmospheres of small sub-Neptunes to reveal potential trends that, beyond equilibrium temperature, could track with planet mass, stellar type, or planetary surface gravity.

\subsection{Observational discriminants of atmospheric origins} \label{sec:nirspec_predictions}

Further observational probes of the atmosphere of GJ 9827 d have the potential to shed more light on its HMMW volatile makeup and on the origins of its volatile-rich atmosphere. In particular, upcoming transit observations with NIRSpec/G395H as part of GO 4098 will open the possibility of detecting carbon species such as CO and CO$_2$, or even sulfur-bearing species (e.g. SO$_2$). The constraining power of NIRISS/SOSS for CO or CO$_2$ abundances is very limited (see Figures \ref{fig:vmr_constraints_allmols}, \ref{fig:vmr_constraints_allmols_clr}, Table \ref{tab:mol_constr}), while both these molecules have larger cross sections and major absorption features in the NIRSpec/G395H bandpass.

A detection of CO$_2$ as the dominant carbon-bearing species would confirm our tentative conclusion of the oxidizing state of the atmosphere and suggest more C than H in the initial silicate endowment of the mantle. However, more detailed modeling would be required to tell apart the origin of the oxygen, either from outgassing from a highly oxidized mantle or direct volatile enrichment from the protoplanetary disk (Figure \ref{fig:comp_HMMWV}). Abundant SO$_2$ is a strong signature of oxidized mantle outgassing, but can also be produced by upper atmosphere photochemistry of H$_2$S and H$_2$O \citep{tsai_photochemically_2023}. Alternatively, if CO dominates, the favored interpretation for the volatile enrichment would be mantle outgassing rather than the accretion of solid-derived volatiles as CO$_2$ would quickly dominate in the most oxidizing conditions. Finally, the detection of abundant methane or ammonia, which are only weakly constrained by our NIRISS/SOSS spectrum, would suggest a carbon-poor silicate mantle (Figure \ref{fig:comp_HMMWV}). 

Further, the atmospheric CO$_2$/CH$_4$ ratio was recently proposed as an observational window into the O/H or H$_2$O/H$_2$ ratio in the deep interior of warm-temperate planets \citep{yang_chemical_2024}. If this finding holds for the $\sim 620\,K$ GJ 9827 d, a detection of carbon-bearing species could help us differentiate between water derived from the accretion of ice-rich solids, or from volatile-poor formation followed by interior atmosphere interactions, although in both cases volatile partitioning between the core, mantle, and atmosphere will alter the atmospheric volatile content. Observations at longer wavelengths could potentially reveal oxygen buildup via the presence of O$_2$ (via the collisionally induced absorption band at $\sim 6\mu$m; \citealt{fauchez_impact_2019}) or O$_3$ at 9.6$\mu$m, but further theoretical work is required to estimate the expected abundance of these species in the atmospheres of small sub-Neptunes such as GJ 9827 d and assess their detectability.

\section{Summary and Conclusions} \label{sec:conclusion}

We obtained two new transit observations of GJ 9827 d with JWST/NIRISS SOSS as part of the JWST sub-Neptune survey (GO 4098). We detect seven candidate flare events in the data, three of which are confirmed as bona fide flares by their time evolution and measured spectral energy distribution. The spectra obtained from both visits are best explained by water features with small $\sim20-40$ppm amplitudes. Our water detection, in agreement with the HST/WFC3 result \citep{roy_water_2023}, is robust even when the TLS effect is considered.

The addition of the JWST/NIRISS SOSS observations enables us to break the degeneracy between low-metallicity cloudy and high mean molecular weight atmospheres, and we detect a high mean molecular weight, metal-enriched and potentially water-rich ``steam world'' atmosphere which from irradiation considerations would be in a well-mixed vapor/supercritical state. The metal-enriched atmosphere scenario is also supported by our and previous nondetections of escaping neutral hydrogen and metastable helium from the atmosphere of GJ 9827 d as escape rates are lower in higher mean molecular weight atmospheres and hydrogen ionization fraction in the escaping flow is larger, eluding observational probes. We estimate, for a pure-H$_2$O envelope on top of a solid rock/iron core, that the hydrosphere makes up between 20 and 40\% of the total planet mass.

The atmospheric metal enrichment could originate from large initial volatile inventories from early accretion, preferential loss of H through mass fractionation, or enrichment through magma ocean-atmosphere interactions over geological timescales. Future observations probing potential absorption from carbon- and sulfur-bearing species could shed more light on which volatile enrichment mechanism(s) shaped the atmosphere of GJ 9827 d.

\vspace{10mm}
\begin{acknowledgments}
We thank the anonymous referee for providing comments that improved our manuscript. This work is based on observations with the NASA/ESA/CSA James Webb Space Telescope, obtained at the Space Telescope Science Institute (STScI) operated by AURA, Inc. All of the data presented in this paper were obtained from the Mikulski Archive for Space Telescopes (MAST) at the Space Telescope Science Institute. The data used in this paper can be found in MAST at the following DOIs: \dataset[10.17909/sx19-8852]{http://dx.doi.org/10.17909/sx19-8852} (\textit{JWST}) and \dataset[10.17909/dvqh-2r48]{http://dx.doi.org/10.17909/dvqh-2r48} (\textit{HST}). The authors thank O. Lim for providing a NIRISS/SOSS throughput file and A. Feinstein for sharing her perspective on stellar flares and their impact on this dataset. C.P.-G. acknowledges support from the NSERC Vanier scholarship and the Trottier Family Foundation. This work was funded by the Institut Trottier de Recherche sur les Exoplanètes (iREx). M.R.\ acknowledges funding from NSERC, Fonds de recherche du Québec – Nature et technologies (FRQNT), and iREx. R.A. acknowledges the SNSF support under the Post-Doc Mobility grant P500PT\_222212 and the support of the Institut Trottier de Recherche sur les Exoplanètes (iREx). M.F.T. acknowledges financial support from the Clarendon Fund and the FRQNT.

\end{acknowledgments}

\clearpage

\appendix
\setcounter{figure}{0}
\renewcommand{\thefigure}{A\arabic{figure}}
\setcounter{table}{0}
\renewcommand{\thetable}{A\arabic{table}}

\section{Comparison of fitted parameters with literature values}\label{sec:comp_lit}

\begin{figure}
    \centering
    \includegraphics[width=0.45\textwidth]{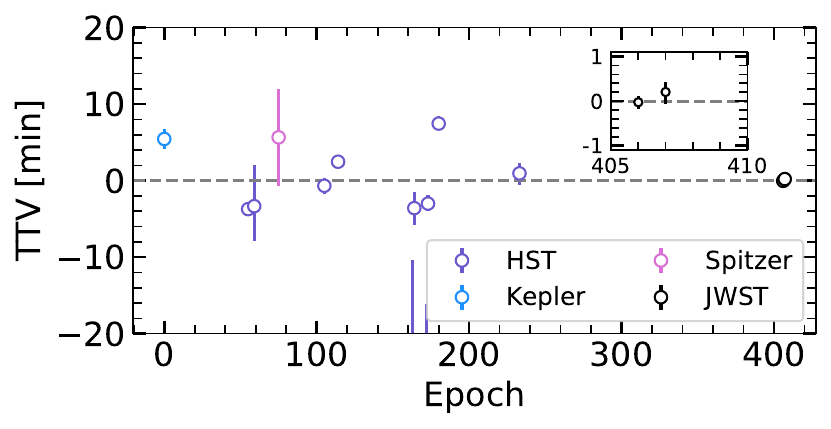}
    \vspace{5mm}
    \vspace{-5mm}\caption{Comparison of the fitted transit time variations for visits 1 and 2 with NIRISS/SOSS (black points) to the previous measurements from Kepler (blue), HST (purple), and Spitzer (pink). We show the transit-timing variations (TTVs) as a function of epoch, after fitting a new ephemeris to all the transit times combined. The inset zooms in on the two NIRISS/SOSS transits presented in this work. The JWST transits nearly double the baseline, occurring over 1000 days after the last HST transit, improving the precision of the orbital period measurement by a factor of five.}
    \label{fig:TTVs}
\end{figure}

Our two measured transit times enable us to refine the ephemeris of GJ 9827 d (see Table \ref{table:pla_star_param}). Not only is the precision of our transit times measurement higher compared to previous sources, but the two NIRISS/SOSS transits also occurred over 1000 days after the previously-measured \textit{HST} transit times (Figure \ref{fig:TTVs}, \citealt{roy_water_2023}).

We do not detect significant transit-timing variations from the two visits, separated by only one planetary orbit. This is in contrast to the 10 measured \textit{HST} transit times, which exhibit TTVs on the order of 5 to 10 minutes (Figure \ref{fig:TTVs}). We note, however, that the \textit{HST} transit times might be biased by the assumed b and a/R$_\star$ fixed in the \textit{HST} light-curve fits to literature values \citep{niraula_three_2017}
which differ significantly from our retrieved impact parameter and semi-major axis (Table \ref{table:wlc_param}). Our fitted orbital parameters and the derived planetary properties agree with literature values \citep{kosiarek_physical_2021,passegger_compact_2024}. Our derived planetary radius has larger uncertainties than the \textit{Spitzer} value even if our $R_p/R_\star$ is more precise, with the error budget dominated by the larger error bars on the stellar radius ($0.58 \pm 0.03 R_\odot$) from \citet{passegger_compact_2024} compared to the isochrone determination from \citet{kosiarek_physical_2021} ($0.579 \pm 0.018 R_\odot$). The more conservative error bars from \citet{passegger_compact_2024} are derived using a joint GP modeling of the transits and radial velocity data with a broad prior informed not only by the isochrone fit to the ESPRESSO-derived stellar parameters but also other determinations using alternative age determination methods.

\begin{table*}
\caption{Fitted planetary parameters to both \textit{JWST}/NIRISS SOSS Order 1+2 transits of GJ 9827d extracted using \texttt{supreme-SPOON} and fitted using \texttt{ExoTEP}. We report the planet-to-star radius ratios separately for the white-light-curve fits to each order. 
}
\centering
\begin{tabular}{lccc}
\hline
\hline
Fitted parameter      & Visit 1 & Visit 2         \\
\hline

Mid-transit time $T_c$ [BJD$_\mathrm{TDB}$]     &  $2460258.90012 \pm 0.00010$ & $2460265.10207_{-0.00019}^{+0.00018}$  \\
Planet-to-star radius ratio $R_p/R_*$:   &    &    \\
~~~~--NIRISS/SOSS order 1   & $0.03157_{-0.00063}^{+0.00062}$  & $0.03126_{-0.00061}^{+0.00070}$      \\
~~~~--NIRISS/SOSS order 2   & $0.03252_{-0.00083}^{+0.00095}$  & $0.03188_{-0.00093}^{+0.00105}$      \\
Semi-major axis $a/R_*$   &   $19.56_{-0.66}^{+0.64}$ & $20.00_{-0.78}^{+0.82}$\\
Impact parameter $b$      &   $0.893 \pm 0.010 $ & $ 0.891_{-0.012}^{+0.011}$ \\

\hline
\label{table:wlc_param}
\end{tabular}
\end{table*}

\section{Consistency of spectrum across binning schemes, data reductions, and instruments}\label{sec:comp_spec}

\begin{figure*}
	\centering
	{
		\includegraphics[width=0.8\textwidth]{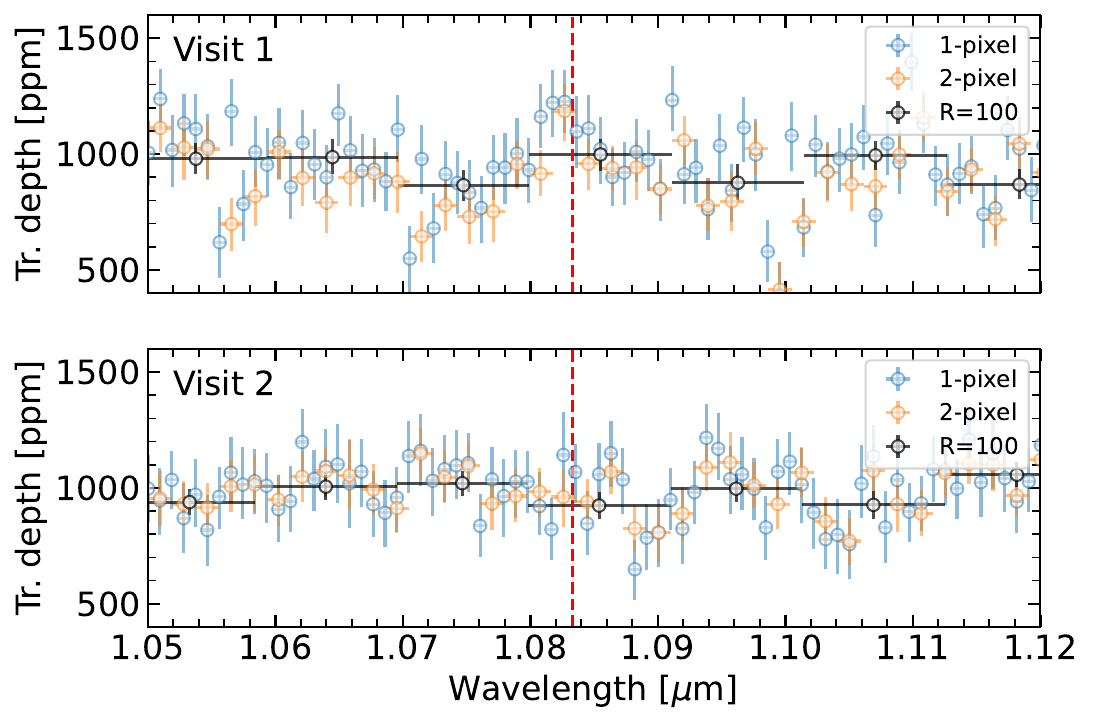}
	}
	\caption{Comparison between the spectra that were extracted at high (1-pixel in blue and 2-pixel in orange) resolution, and the R=100 (black) spectrum, for Visit 1 (top panel) and Visit 2 (bottom panel) using the \texttt{supreme-SPOON} version of the data reduction. The three versions of the transmission spectrum for each visit are consistent to well within their quoted uncertainties in each spectral bin. }
	\label{fig:compare_three_spectra}
\end{figure*}

\begin{figure*}
	\centering
	{
		\includegraphics[width=0.8\textwidth]{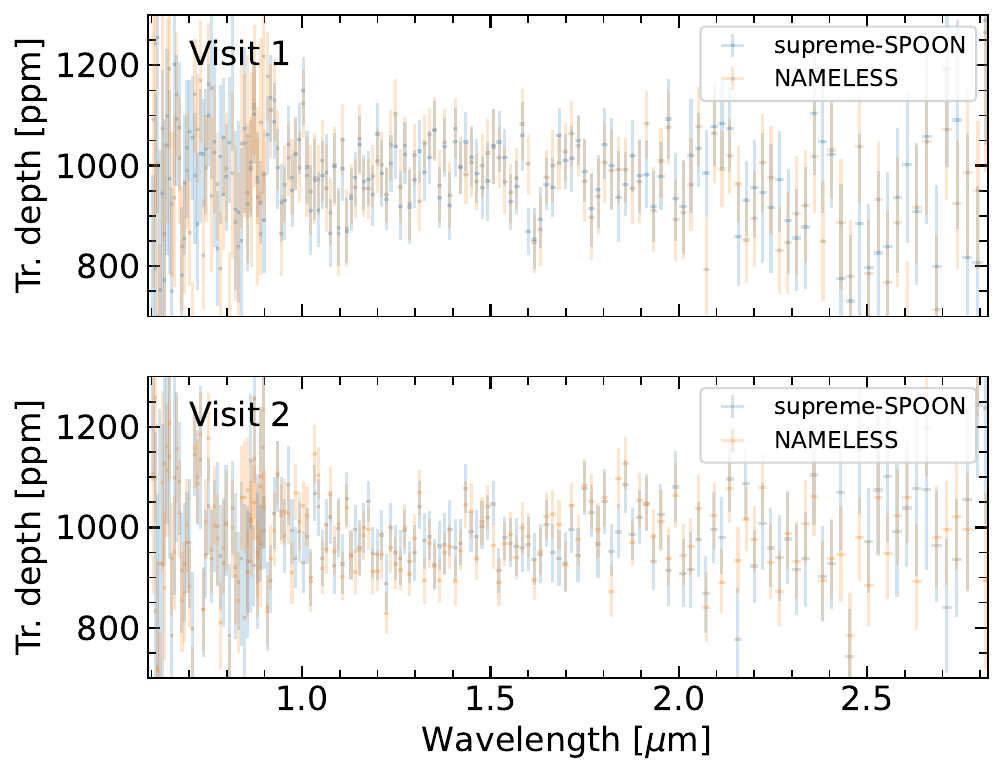}
	}
	\caption{Comparison between the spectra obtained from the \texttt{supreme-SPOON} (blue) and \texttt{NAMELESS} (orange) data reductions. The top panel shows the R=100 spectrum for visit 1, and the bottom panel shows the same thing for visit 2. The spectra agree to better than 1 sigma in almost all spectral bins.}
	\label{fig:compare_reductions}
\end{figure*}

\begin{figure}
    \centering
    \includegraphics[width=0.8\textwidth]{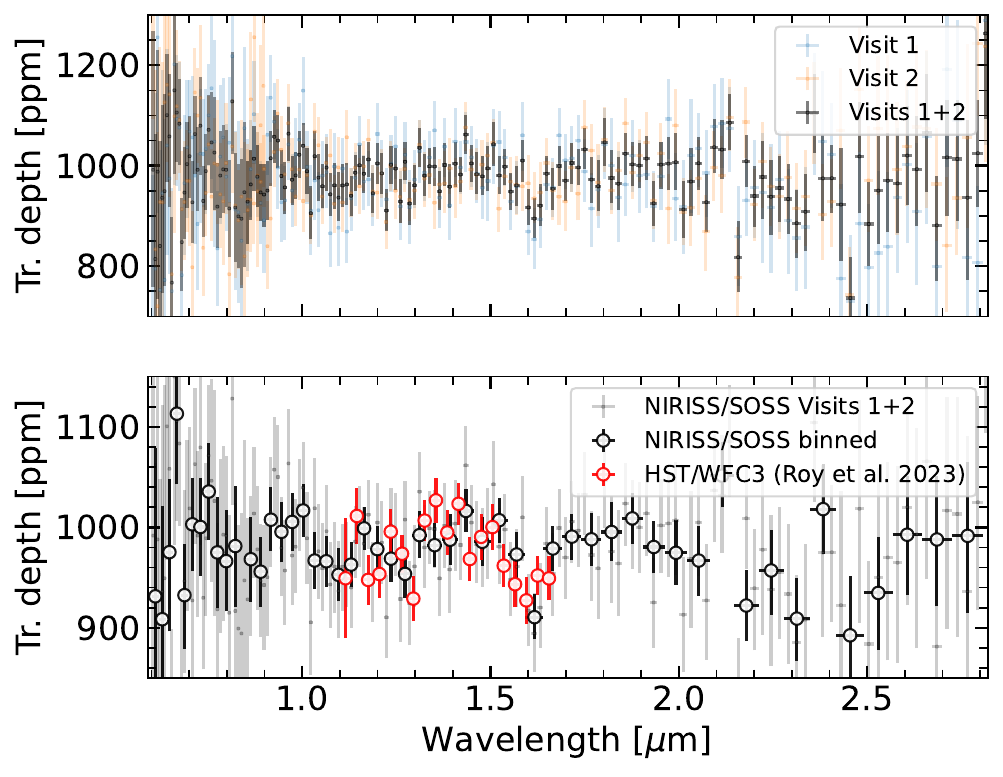}
    \vspace{5mm}
    \vspace{-5mm}\caption{Comparison of the spectrum of GJ 9827 d over both JWST/NIRISS SOSS transit epochs and with the HST/WFC3 observations. \textit{Top panel:} Transmission of GJ 9827 d as observed during the first (blue) and second (orange) visit, with the spectrum of visit 1 shifted down by about 40 ppm to match the median value of the visit 2 spectrum. The combined spectrum from both visits is shown in black. \textit{Bottom panel: }Combined JWST/NIRISS SOSS transmission spectrum (gray) along with binned version (black) and the spectrum as observed from 10 HST/WFC3 transits \citep{roy_water_2023}, shifted up by 10 ppm.}
    \label{fig:pipeline_comparison}
\end{figure}

We compare for each visit the spectra extracted 1- and 2-pixel resolution with the R=100 spectrum for the \texttt{supreme-SPOON} reduction and find that they agree well within 1$\sigma$ over the wavelength range they share in common (Figure \ref{fig:compare_three_spectra}). We also compare the spectra obtained from the \texttt{supreme-SPOON} and NAMELESS reductions for each visit, and find a similarly satisfactory agreement (Figure \ref{fig:compare_reductions}). Finally, we compare the \textit{JWST} spectrum to the \textit{HST}/WFC3 transmission spectrum of GJ 9827d and find that they match well over the bandpass of \textit{HST}/WFC3, except for a small vertical shift (Figure \ref{fig:pipeline_comparison}). The transmission spectrum is provided in Table \ref{tab:trans_sp} and in a machine-readable format along with the published article.

\section{Analysis of candidate stellar flares}\label{ssec:intransit_varia}
The multiple long and short timescale events in the observed light-curves can naturally be explained by stellar activity, which principally consists of stellar flares or spot-crossing events. Both events will display different temporal and spectral variations and should be distinguishable in our data. On one hand, the red optical/NIR spectrum of flares evolves on timescales of $\sim$10 s. It is composed of a rapid period of prompt emission with an effective temperature ($T_\mathrm{eff}$) of $\sim$9000 K followed by a more gradual period of emission with a $T_\mathrm{eff}$ of $\sim$5000 K \citep{kowalski_2016}. As a result, the observed $T_\mathrm{eff}$ of the post-subtraction stellar spectrum rapidly increases and then slowly decays during a flare and follows a fast-rise, exponential decay (FRED) temporal profile. Conversely, variations in the residual spectrum resulting from a spot-crossing event occur more slowly and with a symmetrical temporal profile well-described by a Gaussian (e.g., \citealt{beky_2014, schutte_2023, biagiotti_2024}). Furthermore, a flare spectrum is brightest at near-UV and blue optical wavelengths (0.2--0.5 $\mu$m), placing the NIRISS/SOSS wavelength range in the Rayleigh-Jeans tail during the flare peak. Meanwhile, starspot spectra are generally brighter at IR than optical wavelengths \citep{waalkes_2024, biagiotti_2024}. Although it only enables us to sparsely sample the potentially rapidly-evolving brightness profiles of flares, the 16.5-s cadence of the NIRISS/SOSS observations is deemed sufficient to determine whether the $T_\mathrm{eff}$ evolution favors the interpretation of a mid-transit event as either a spot crossing event or flare through the maximum $T_\mathrm{eff}$ value and the presence of (a)symmetry in the $T_\mathrm{eff}$ evolution.



We flux-calibrate the extracted spectra from the \texttt{supreme-SPOON} pipeline against a K7V PHOENIX spectrum obtained from the MUSCLES Extension archive \citep{behr_2023}, and follow the procedure applied to NIRISS/SOSS observations of TRAPPIST-1 \citep{howard_characterizing_2023}. The K7V PHOENIX spectrum is computed from the Lyon BT-Settl CIFIST 2011\_2015 grid \citep{allard_2016} for a star of $T_\mathrm{eff}$=4124 K, $d_*$=86.7$\pm$0.3 pc, $M_*$=0.72$\pm$0.03 M$_\odot$, and $R_*$=0.67$\pm$0.01 R$_\odot$, and scaled to match the distance to GJ 9827. Comparison with a blackbody spectrum at the effective temperature of GJ 9827 reveals that the flux in the atmosphere model exceeds the blackbody flux by a factor of 1.348 in the wavelength range of 0.6--1.6 $\mu$m where the majority of flare emission occurs in NIRISS observations from the shape of the flare blackbody spectrum. As a result, we further scale the PHOENIX spectrum by a factor of 0.742 to maximize calibration accuracy within this region. The color correction for each visit is obtained by taking the ratio between the PHOENIX model spectrum and an average spectrum of all quiescent integrations during the visit. For the construction of this quiescent spectrum, we select integrations that are both out of transit and in time intervals where the stellar activity (assessed from the NIRISS/SOSS light-curve integrated in the TESS bandpass) is low. Before computing the ratio of the observed and model spectra, we bin the spectrum in resolution (200 columns per wavelength bin for order 1 and 160 for order 2) in order to remove localized spectral features \citep{howard_characterizing_2023}. This ratio is then used to flux-calibrate each of the extracted spectra in the time series and to obtain a flux-calibrated quiescent spectrum. 

\begin{figure*}
	\centering
        \subfigure
	{
		\includegraphics[trim=0 0 0 0, width=0.98\textwidth]{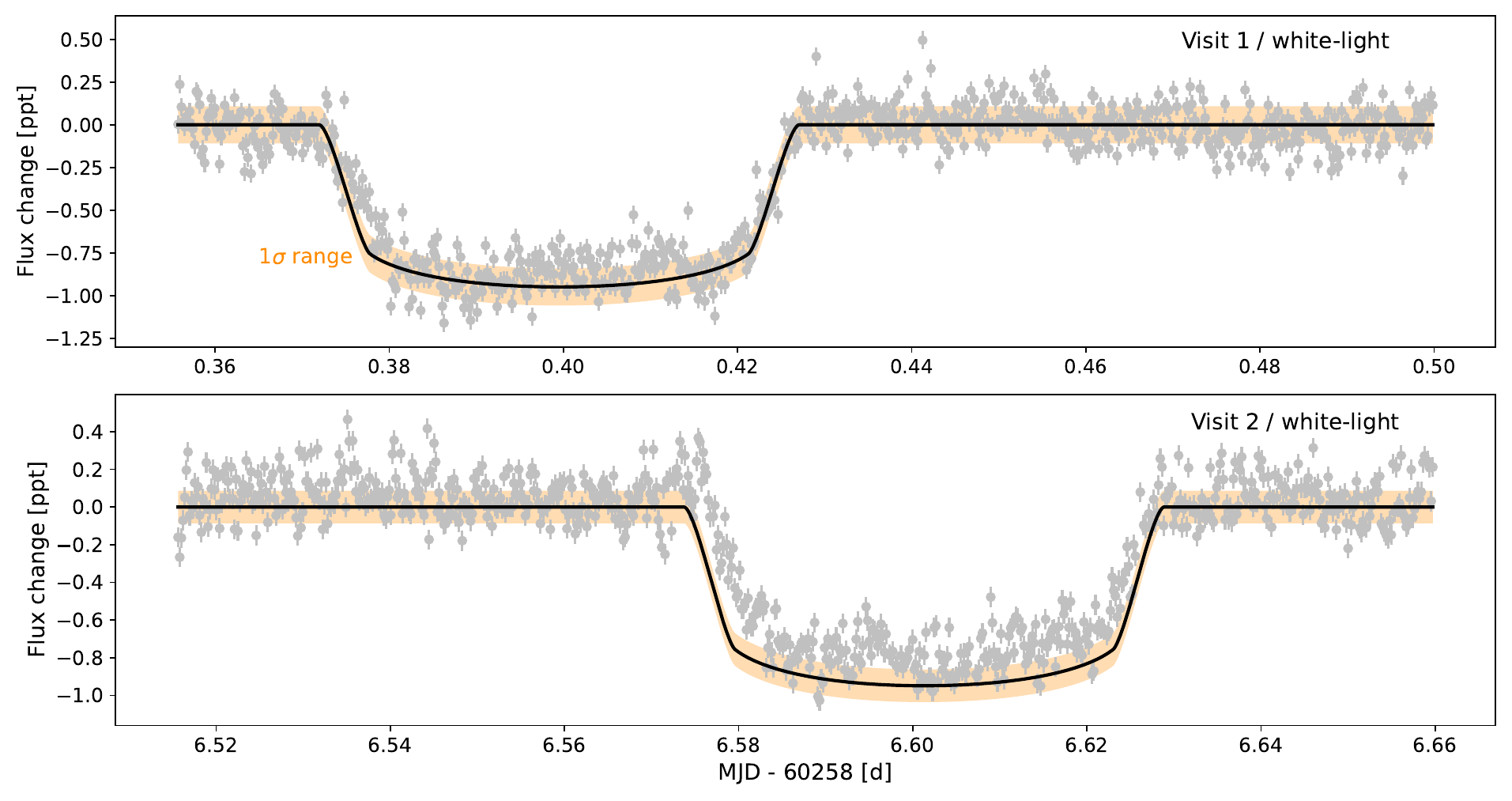}
	}
        \subfigure
	{
		\includegraphics[trim=20 15 14 0, width=0.48\textwidth]{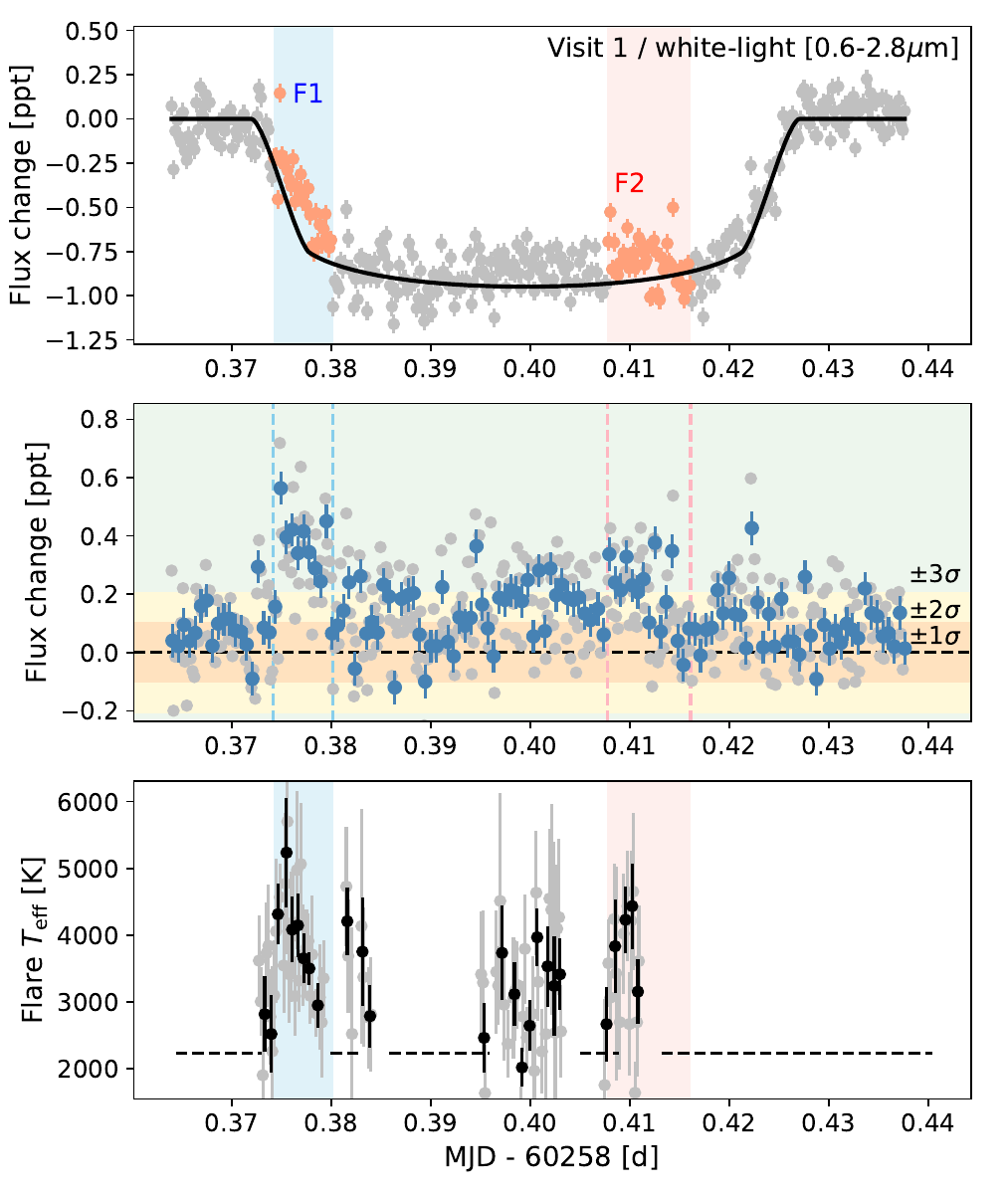}
	}
        \subfigure
	{
		\includegraphics[trim=14 15 20 0, width=0.48\textwidth]{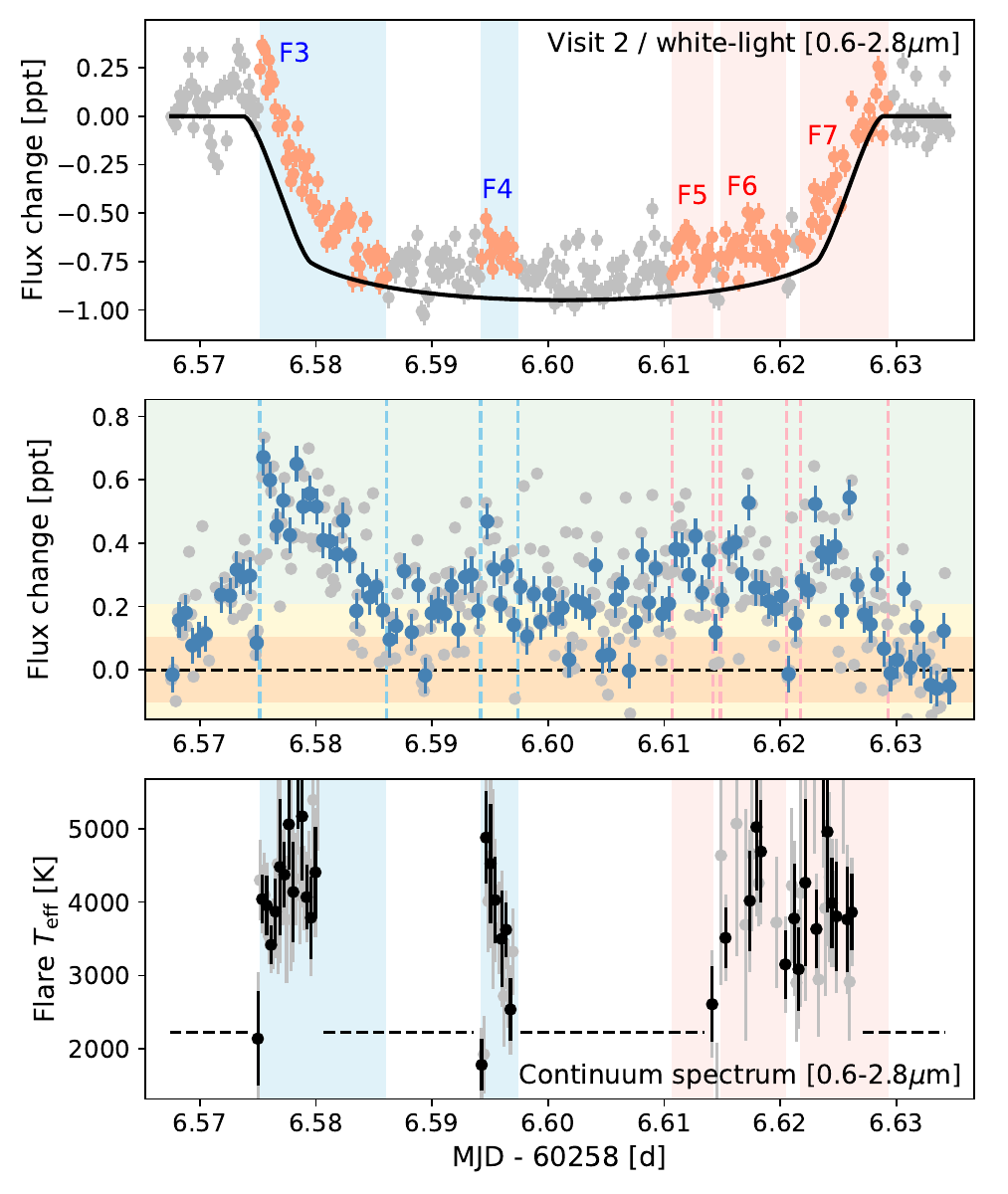}
	}
	\caption{\textit{Top panels}: WL light-curves showing the full baseline versus the transit model. The 1$\sigma$ flux range during times of quiescence is highlighted. We opt to use the transit depth from visit 1 during visit 2 due to degeneracy between the depth and the flare decay tail of F3. \textit{Bottom left panels}: The WL transit light-curve, TESS band time series, and $T_\mathrm{eff}$ time series for the flare-only continuum spectra are shown for visit 1. Times during flares are highlighted, where flares are defined as events with TESS band flux increases with two or more points exceeding the local noise by 3$\sigma$ in the binned light-curve. The 16.5-s cadence data are shown in gray, while binned time series are shown in blue and black for the TESS and $T_\mathrm{eff}$ time series, respectively. A fitted linear trend in the out-of-transit light-curve has been removed from the WL light-curve to better show flare events. The transit model shown in the top panel has been subtracted from the TESS time series. Gaps in the $T_\mathrm{eff}$ time series are due to low S/N or nondetections outside of the flare peak. \textit{Bottom right panels:} Same as the bottom left panels, but for visit 2.}
	\label{fig:flare_timeseries}
\end{figure*}

We construct flare-only spectra by calculating the difference between each flux-calibrated spectrum in the time series (normalized by the transit light-curve), and the quiescent spectrum. To obtain the \texttt{batman} model shown in Figure \ref{fig:flare_timeseries}, we choose to fix the transit depth in the model fitted to the visit 2 light-curve to that obtained from a fit to the visit 1 transit because of the long decay tail of the largest F3 flare. This causes a difference in apparent ``fit quality'' with the white light curve fit used to obtain the transmission spectrum (Figure \ref{fig:wlc_slc_fit}) but ensures that the flare analysis is not biased by to degeneracy between the flare tail and transit depth. We carefully select the lowest-activity epochs when computing the median out of transit flux.
The measured flare properties are more sensitive to the inclusion of small flux excursions in the quiescent baseline than are the properties of a time-averaged transmission spectrum, requiring a separate \texttt{batman} model subtraction.

We fit the $T_\mathrm{eff}$ of the flare-only spectrum for each integration using a single-temperature Planck function model \citep{howard_characterizing_2023}, allowing us to obtain a flare effective temperature time series for each visit (Figure \ref{fig:flare_timeseries}. We also use the convolution of each NIRISS/SOSS spectrum with the transmission curve of TESS in the 0.6--1.0 $\mu$m range, to construct a light-curve that can be compared with the light-curves of thousands of K and M dwarf flares observed at a comparable 20-s cadence by the TESS mission (e.g. \citealt{gilbert_2022, howard_macgregor_2022, paudel_2024}).

We flag as flares events in the post-processed TESS band light-curve where at least two consecutive integrations exceed the measured visit-specific scatter by $\geq3\sigma$. Flare energies are calculated using the quiescent energy of GJ 9827 in the TESS bandpass ($Q_0$=1.000$\pm$0.002$\times$10$^{32}$ erg s$^{-1}$) and the flare equivalent duration (ED; \citealt{howard_characterizing_2023}). We detect seven small candidate flare events with TESS band energies ranging from 7.40$\pm$0.60$\times$10$^{30}$ to 3.85$\pm$0.12$\times$10$^{31}$ erg across both visits (labeled F1--F7; see Figure \ref{fig:flare_timeseries}). The presence of seven flares with $E_\mathrm{TESS}\geq$7.4$\times$10$^{30}$erg during a total stare time of 0.289 d qualifies GJ 9827 as an active K-dwarf, whose flare rate is typical of other active field stars of spectral class K7--M0 \citep{howard_2019}. 

\begin{figure*}
	\centering
	{
		\includegraphics[width=0.99\textwidth]{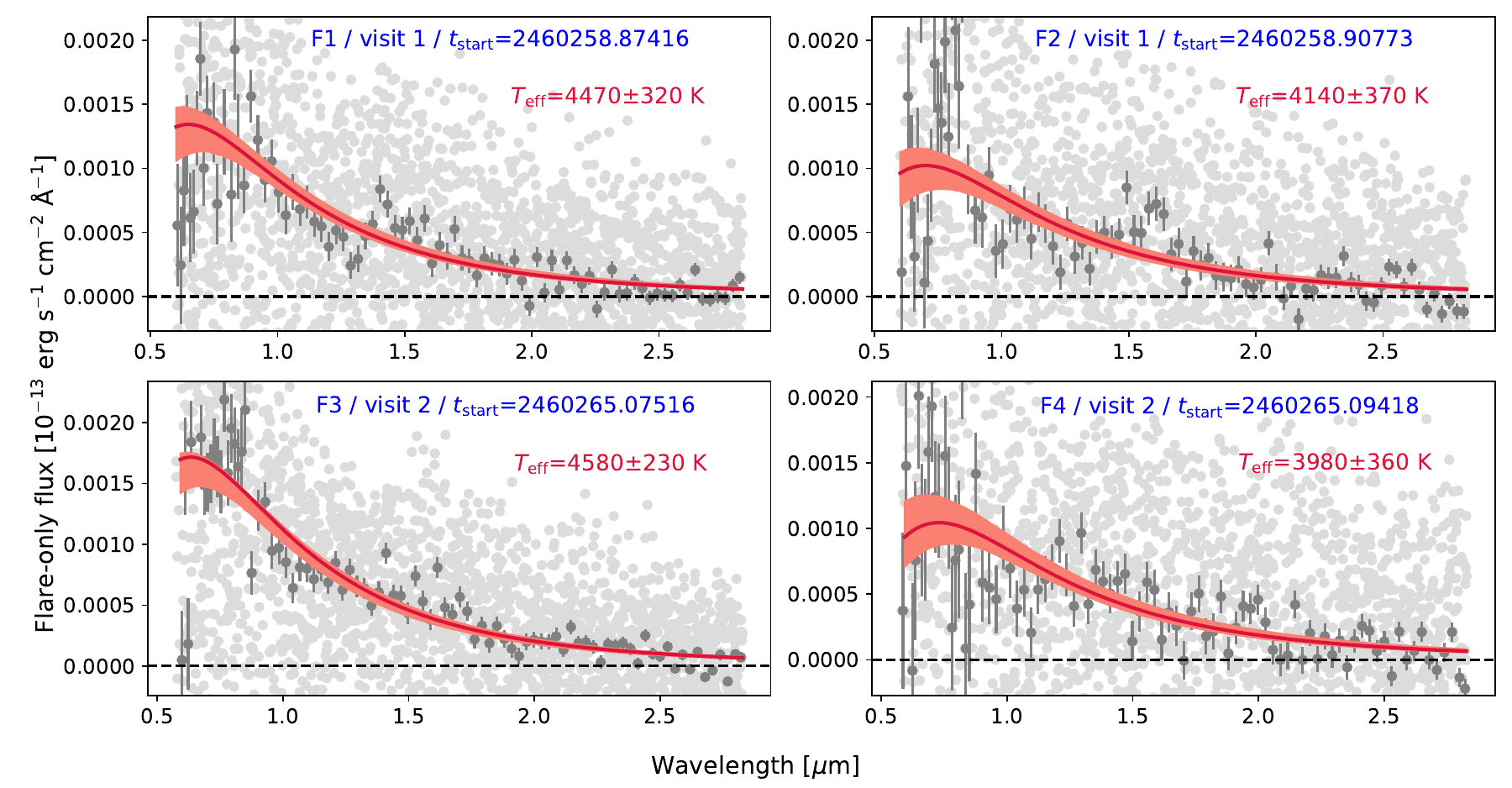}
	}
	\caption{Time-averaged spectra of all integrations during the flare peak times of the F1--F4 candidates. Each spectrum is obtained by subtracting the out-of-flare stellar spectrum to create a flare-only spectrum. The flare-only spectrum is binned in wavelength and fit with a single-temperature blackbody model to obtain the hottest $T_\mathrm{eff}$ reached during each event.}
	\label{fig:example_flare_spectra}
\end{figure*}

We measure the $T_\mathrm{eff}$ values from the flare-only spectra of each event, averaging over all integrations during the peak times of the flare to increase the signal to noise. The S/N of the $T_\mathrm{eff}$ time series of the F1, F3, and F4 events are sufficient to observe the rapid rise and slow decay profile typical of stellar flares, but not starspots. For example, the rise time of the F1, F3, and F4 $T_\mathrm{eff}$ time series lasts 2.17, 0.29, and 0.27 min, respectively. The rise times of spot crossing events during transits of HAT-P-18 b and HATS-2 b are 6.9, 11.2, and 13.6 min, respectively \citep{fournier-tondreau_near-infrared_2023, biagiotti_2024}. The $T_\mathrm{eff}$ value obtained from the flare-only spectrum averaged over the peak integrations for each of these same events is 4470$\pm$320, 4580$\pm$230, and 3980$\pm$360 K, respectively (Figure \ref{fig:example_flare_spectra}). 

The peak continuum temperatures are similar to those of four flares from the M8 dwarf TRAPPIST-1 observed with NIRISS and NIRSpec. Although for GJ 9827 these temperatures are close to the effective temperature of the star, the flux amplitude of the flare-only spectrum used to measure the $T_\mathrm{eff}$ values is $\geq$5 times larger than the uncertainty in the subtraction of the quiescent stellar spectrum from 0.6--1.7~$\mu$m, and $\geq$50 larger than the uncertainty at 0.9~$\mu$m. The TRAPPIST-1 flares also display a FRED time series in $T_\mathrm{eff}$ that peaks at values of up to 5300 K \citep{howard_characterizing_2023}. We fit an empirical stellar flare template model \citep{tovar-mendoza_2022} and a Gaussian model to the $T_\mathrm{eff}$ time series of flares F1, F3, and F4 and measure the residual sum of squares (RSS) for each model to quantify the degree of asymmetry. The flare model is preferred in each case by a factor of 3.1, $>$5, and 4.1 for F1, F3, and F4, respectively. We conclude that the combination of the extremely rapid and asymmetrical continuum temperature evolution and the 4000--5000 K peak $T_\mathrm{eff}$ values of these events are best explained by the flare interpretation.

While for TRAPPIST-1, the H$\alpha$ line exhibits variations that track the time evolution of the flare, we do not find clear flare evidence in H$\alpha$ for GJ 9827. Although the TESS-band flare energies we obtain for GJ 9827 are similar to those of TRAPPIST-1, a quiescent K7V star is three orders of magnitude brighter at 0.656~$\mu$m than an M7.5V star, such that a flare from GJ 9827 must be 10$^3$ times more luminous in the H$\alpha$ line to produce the same H$\alpha$ light-curve as a flare from TRAPPIST-1. The TRAPPIST-1 flares emitted only 10\% as much energy in the H$\alpha$ line as in the continuum. If this energy distribution holds for late K dwarfs, the presence of continuum enhancements is, therefore, more indicative of flaring than are H$\alpha$ enhancements, which could explain our nondetection.

We use optical-NUV and NUV-FUV flare energy scaling relations to translate our measured flare energies to FUV (0.09--0.17 $\mu$m) and NUV (0.17--0.32 $\mu$m) energies (Table \ref{tab:flares_data}). 
Even small flares observed at NIR wavelengths can produce significant enhancements in the stellar emission at UV wavelengths due to the peak of the flare SED from 0.2--0.4 $\mu$m \citep{jackman_2023, paudel_2024}. We employ the $E_\mathrm{TESS,FWHM}$-$E_\mathrm{NUV,tot}$ relation obtained from simultaneous 20-s cadence TESS and \textit{Swift} observations to estimate the total NUV energy emitted during the F1--F7 flares (Howard et al., in preparation). This relation leverages the time resolution of the 20-s cadence TESS data to measure the energy emitted in the TESS band with a two-fold increase in precision. We propagate the mean and 1$\sigma$ uncertainty in the estimated NUV energies into the FUV using the ratio of flare emission at FUV to NUV wavelengths. This ratio, $\mathcal{R}_\mathrm{FUV/NUV}=0.49\pm0.02$, was recently measured from GALEX observations of a sample of 182 flares from 158 field stars which provide a strong constraint on the FUV emission of any NUV flare \citep{berger_2023}. 

The combined FUV and NUV energies of F1--F7 make up 32\% and 3\% of the integrated quiescent stellar FUV and NUV stellar emission throughout all transit observations, respectively. The flares can contribute larger fractions of the total UV irradiation over shorter time intervals. For example, the F3 event contributes 82\% and 8\% of the integrated quiescent emission at FUV and NUV wavelengths respectively during the first hour of the transit observation during visit 2. Variable stellar emission at FUV and NUV wavelengths can impact the abundances of volatile species in the upper atmosphere through photochemical processes, and the frequent flares we observe suggest that photodissociation might be even more prevalent (especially during the second NIRISS/SOSS transit) than estimated from observations of the quiet state of GJ 9827. 

\section{Modeling of the transit light source effect}\label{sec:stellar_ctm}

If the stellar surface is heterogeneous, in the presence of dark spots or bright faculae, the average stellar surface spectrum may not be a good representation of the spectrum of the star along the transit chord (see e.g. \citealt{sing_hubble_2011,berta_gj1214_2011,rackham_transit_2018}). In particular, for the case of GJ 9827 d, unocculted stellar spots can mimic water features in the transmission spectra of planets around cool stars (e.g., \citealt{barclay_stellar_2021,brande_mirage_2022}).

In that case, the transmission spectrum of the planet is multiplied by a wavelength-dependent factor $\epsilon_{\lambda, \, \rm{het}}$ \citep{rackham_transit_2018, pinhas_retrieval_2018}:
\begin{equation}
    \epsilon_{\lambda, \, \rm{het}} = \frac{1}{1 - \displaystyle\sum_{i=1}^{N_{\rm{het}}} f_{\mathrm{het}, \, i} \left(1 - \frac{I_{\lambda, \, \mathrm{het}, \, i} \left(T_\mathrm{het, i}\right)}{I_{\lambda, \, \rm{phot}}\left(T_\mathrm{phot}\right)} \right)}
\label{eq:stellar_contam_factor}   
\end{equation}
where $f_{\mathrm{het}, \, i}$ describes the fraction of the stellar disk covered by heterogeneity $i$, while $I_{\lambda, \, \mathrm{het}, \, i}$, and $I_{\lambda, \, \rm{phot}}$ correspond respectively to the specific intensity of the stellar heterogeneity $i$ with temperature $T_\mathrm{het, i}$ and of the stellar photosphere with temperature $T_\mathrm{phot}$. 

In our modeling, we assume one or two populations of stellar heterogeneities: 
\begin{enumerate}
    \item A single population of stellar spots colder than the stellar photosphere: $N_{\rm{het}}=1$ in Eq. \ref{eq:stellar_contam_factor} and $T_{\rm{spot}}<T_{\rm{phot}}$;
    \item Two populations of heterogeneities ($N_{\rm{het}}=2$ in Eq. \ref{eq:stellar_contam_factor}) corresponding to stellar spots ($T_{\rm{spot}}<T_{\rm{phot}}$) and faculae ($T_{\rm{fac}}>T_{\rm{phot}}$), respectively.
\end{enumerate}

In our modeling, the spot temperature is parameterized as $T_\mathrm{spot} = T_\mathrm{phot} + \Delta T_\mathrm{spot}$, with $\Delta T_\mathrm{spot}<0$, and $\Delta T_\mathrm{fac}>0$ similarly describes the temperature of the faculae component in the two-heterogeneities case.

\begin{figure*}
	\centering
	{
		\includegraphics[width=0.8\textwidth]{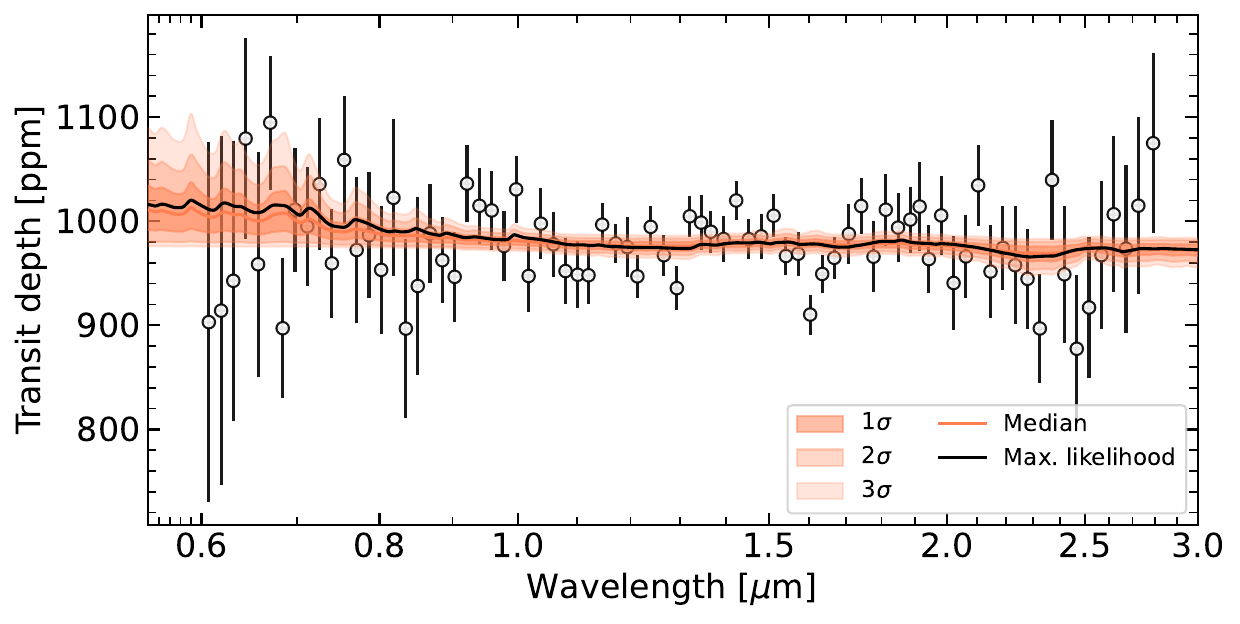}
	}
	\caption{Result from the \texttt{stctm} fit (spots only) to the NIRISS/SOSS + HST/WFC3 spectrum of GJ 9827 d. The combined JWST+HST spectrum is binned to R=50 for visual purposes (black points). The 1, 2, and 3$\sigma$ constraints on the retrieved stellar contamination-only spectra are shown (semi-transparent regions) as well as the median (coral line) and best-fit spectrum (black line). Stellar contamination without a planetary atmosphere signal cannot reproduce the water features observed in the transmission spectrum of GJ 9827 d (see Figure \ref{fig:retrieval_spectrum_fit} for an atmosphere+stellar contamination fit).}
	\label{fig:stctm_retrieved_spectrum}
\end{figure*}

We display the retrieved distribution of stellar contamination-only models that provide the best match to the NIRISS/SOSS + HST/WFC3 spectrum of GJ 9827 d in the spots-only case, and find that the observed water features are inconsistent with stellar contamination alone (Figure \ref{fig:stctm_retrieved_spectrum}). 



\begin{table*}
\centering
\begin{tabular}{ccccc}
\hline
\hline
Instrument               &  Wavelength            &  Depth       & +1$\sigma$ &  -1$\sigma$ \\
                         &  [$\mu$m]              &  [ppm]       & [ppm]      &  [ppm] \\
\hline
  NIRISS/SOSS order 2  &  0.601 -- 0.607    &  991.8     & 275.6    &  232.4 \\
  NIRISS/SOSS order 2  &  0.607 -- 0.613    &  814.2     & 249.8    &  255.4 \\
  NIRISS/SOSS order 2  &  0.613 -- 0.619    &  988.0     & 242.3    &  253.4 \\
  NIRISS/SOSS order 2  &  ...    &  ...     & ...    &  ... \\

  NIRISS/SOSS order 2  &  0.826 -- 0.835    &  899.2     & 103.5    &  102.3 \\
  NIRISS/SOSS order 2  &  0.835 -- 0.843    &  894.3     & 119.6    &  138.8 \\
  NIRISS/SOSS order 2  &  0.843 -- 0.850    &  947.1     & 139.4    &  148.9 \\
\hline
 NIRISS/SOSS order 1 &  0.851 -- 0.860    &  928.3     & 81.4    &  84.2 \\
               NIRISS/SOSS order 1 &  0.860 -- 0.867    &  962.7     & 68.7    &  70.4 \\
               NIRISS/SOSS order 1 &  0.867 -- 0.876    &  1013.7     & 66.2    &  64.1 \\
               NIRISS/SOSS order 1 &  0.876 -- 0.885    &  978.1     & 63.4    &  60.0 \\
               NIRISS/SOSS order 1 &  0.885 -- 0.894    &  946.8     & 56.6    &  56.2 \\
               NIRISS/SOSS order 1 &  0.894 -- 0.903    &  942.6     & 59.9    &  64.5 \\

               NIRISS/SOSS order 1 & ...    &  ...     & ...   &  ... \\
               NIRISS/SOSS order 1 &  2.645 -- 2.672    &  1065.9     & 115.8    &  117.9 \\
NIRISS/SOSS order 1 &  2.672 -- 2.698    &  881.3     & 98.3    &  110.2 \\
               NIRISS/SOSS order 1 &  2.698 -- 2.725    &  1016.5     & 113.4    &  117.4 \\
               NIRISS/SOSS order 1 &  2.725 -- 2.753    &  1013.3     & 116.6    &  122.5 \\
               NIRISS/SOSS order 1 &  2.753 -- 2.779    &  936.1     & 155.2    &  163.7 \\
               NIRISS/SOSS order 1 &  2.779 -- 2.808    &  1025.1     & 105.6    &  107.7 \\
               NIRISS/SOSS order 1 &  2.808 -- 2.820    &  1263.4     & 176.9    &  169.8 \\
\hline
\hline
\end{tabular}
\caption{\label{tab:trans_sp} Truncated NIRISS/SOSS spectrum of GJ 9827 d, using the \texttt{supreme-SPOON} reduction, fitted at R=100. A full version of this spectrum is provided as a machine-readable table in the electronic journal version.}
\end{table*}

\begin{table*}
\centering
\caption{Properties of flare candidates from GJ 9827 during visits 1 and 2. Columns are flare ID, flare start and stop time, peak significance of detection in the binned TESS band light-curve, TESS band flare energy, estimated NUV flare energy, estimated FUV energy, and $T_\mathrm{eff}$ of the flare-only spectrum at peak. The NUV and FUV energies are derived from optical-NUV scaling relations based on simultaneous Swift and TESS flare observations in \citet{paudel_2024} and Howard et al. (in prep.), and GALEX flare observations in \citet{berger_2023}.}
\begin{tabular}{cccccccc}
\hline
\hline
 Flare ID & $t_\mathrm{start}$ & $t_\mathrm{end}$ & $\sigma_\mathrm{flare}$ & $E_\mathrm{TESS}$ & $E_\mathrm{NUV}$ & $E_\mathrm{FUV}$ & $T_\mathrm{eff}$ \\
 & BJD & BJD & & 10$^{30}$ erg & 10$^{30}$ erg & 10$^{30}$ erg & K \\
\hline
F1 & 2460258.87416 & 2460258.88015 & 5.4 & 13.5$\pm$0.9 & 8.37$\pm$0.81 & 4.1$\pm$0.53 & 4470$\pm$320 \\
F2 & 2460258.90773 & 2460258.91609 & 3.2 & 10.9$\pm$1.0 & 5.06$\pm$0.67 & 2.48$\pm$0.37 & 4140$\pm$370 \\
F3 & 2460265.07516 & 2460265.08607 & 6.8 & 38.5$\pm$1.2 & 21.28$\pm$0.94 & 10.41$\pm$0.92 & 4580$\pm$230 \\
F4 & 2460265.09418 & 2460265.09741 & 4.7 & 7.4$\pm$0.6 & 5.61$\pm$0.68 & 2.75$\pm$0.4 & 3980$\pm$360 \\
F5 & 2460265.11065 & 2460265.11423 & 3.9 & 11.0$\pm$0.7 & 5.52$\pm$0.64 & 2.71$\pm$0.39 & 5500$\pm$700 \\
F6 & 2460265.11484 & 2460265.12051 & 4.1 & 14.8$\pm$0.9 & 6.6$\pm$0.74 & 3.23$\pm$0.45 & 5800$\pm$770 \\
F7 & 2460265.12171 & 2460265.1293 & 5.3 & 17.4$\pm$1.0 & 10.32$\pm$0.91 & 5.04$\pm$0.6 & 4520$\pm$360 \\
\hline
\hline
\end{tabular}
\label{tab:flares_data}
\end{table*}

\section{CHAIN atmospheric escape modeling setup}\label{sec:chain_description}

The CHAIN model combines a 1D hydrodynamic upper atmosphere model based on that of \citet{kubyshkina_grid_2018} solving the hydrodynamic equations with the non-local thermodynamical equilibrium (NLTE) photoionization and radiative transfer code \texttt{Cloudy} \citep{ferland2017RMxAA..53..385F} providing detailed (photo)chemistry calculations, hence, realistic heating and cooling rates. The Cloudy code can account for the 30 lightest chemical elements from hydrogen to zinc and supports various compositions (namely, abundances of individual elements (He-Zn) can be set to zero or scaled relative to hydrogen); the code can also account for a range of common molecular species, which are typically present below the 1~$\mu$bar pressure level and destroyed in photodissociation reactions at higher altitudes (e.g., \citealp{egger_unveiling_2024}). 
The hydrodynamic part of the CHAIN code assumes that the atmosphere is well-mixed within the simulation domain (for close-in planets typically set between 1--100\,mbar level and 1--2 Roche radii of the planet) and treats it as a uniform outflow with the mean molecular weight adjusted to reproduce the employed composition. To verify this assumption, the code controls that the exobase position is above the sonic point throughout the simulation. 

The code accounts for the full stellar energy distribution between infrared and X-ray. As a proxy for GJ~9827, we employ the shape of the spectrum of WASP-43 provided by the Measurements of the Ultraviolet Spectral Characteristics of Low-mass Exoplanetary Systems (MUSCLES) Treasury Survey (version 23, \citealt{france2016ApJ...820...89F,Youngblood2016ApJ...824..101Y,Loyd2016ApJ...824..102L}. WASP-43 b has a mass, temperature, and luminosity similar to those of GJ~9827. We further scale the visible-infrared part of the spectrum to reproduce the bolometric luminosity of GJ~9827 and the X-ray plus extreme ultraviolet part (XUV) and anchor our results in the previous literature estimate \citep{carleo_multiwavelength_2021}, who calculate the XUV flux received by GJ~9827~d of $\sim4300$~${\rm erg\,s^{-1}cm^{-2}}$ using the method outlined in \citet{Linsky2014ApJ...780...61L}. This value could potentially be larger during flare events (Section \ref{ssec:intransit_varia}). 
We further verified that the shape of the spectrum does not have a crucial effect on our results by re-running the simulations employing Sun-like and GJ~436 spectra; we found that it affects to some extent the abundances of ion species at low altitudes but not the escape rates. The overall effects are similar to those described in \citet{kubyshkina2024A&A...684A..26K}. We note that CHAIN does not employ a multi-fluid approach, and we rely on the estimations of the maximum mass of species dragged by the hydrodynamic outflow to \citep{hunten1987fractionation}, which may suffer from some inconsistencies \citep{Linssen2024high_metallicities}. Therefore, it is not clear at this stage whether the oxygen ions present at high altitudes would escape along with the hydrogen.

\section{Model comparison and posterior distributions from atmosphere retrievals}

\begin{figure*}
    \centering
    \includegraphics[width=0.85\textwidth]{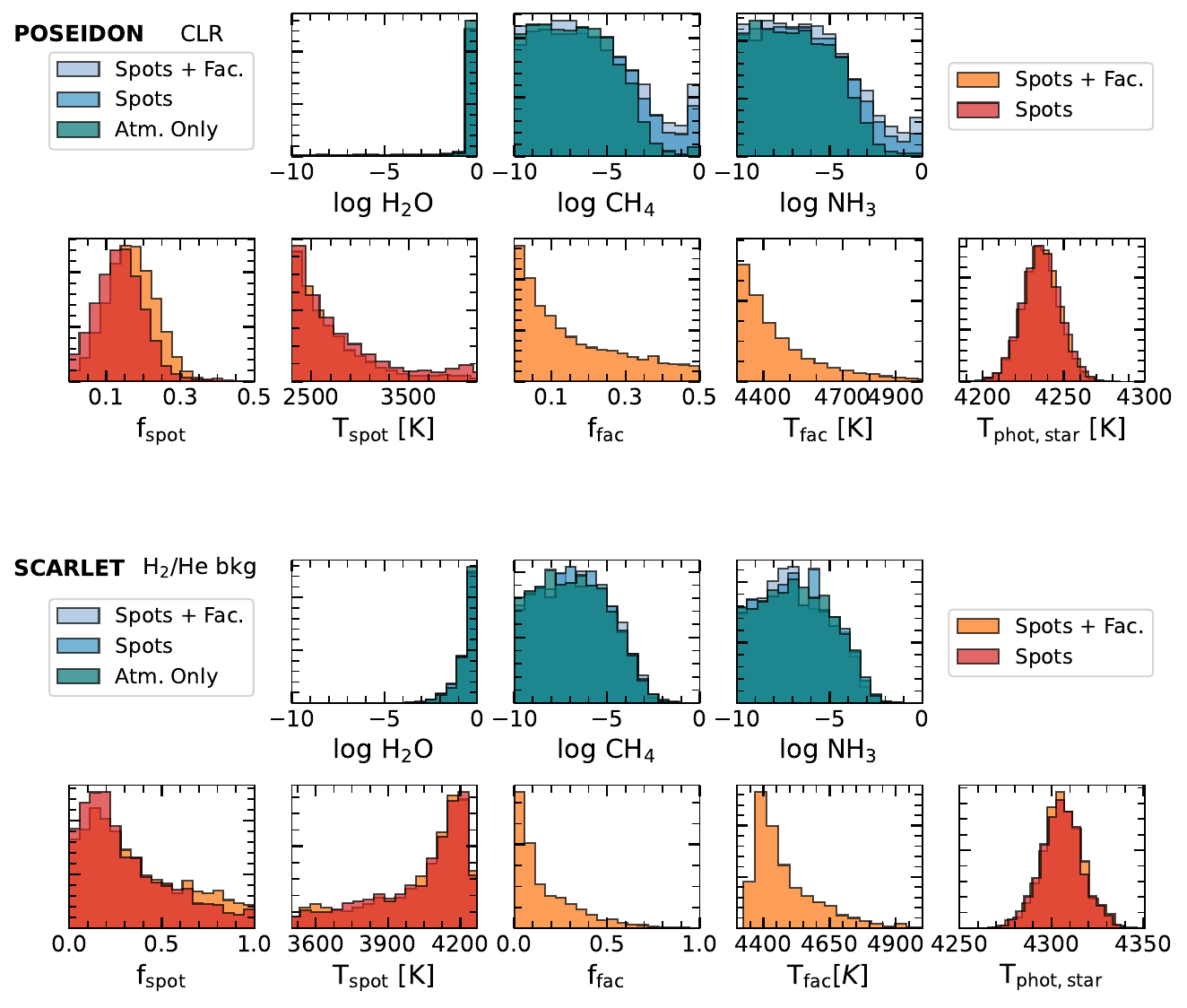}
    \vspace{5mm}
    \vspace{-5mm}\caption{Posterior distributions for important atmospheric volatile species (top, blue) and the stellar heterogeneity component (bottom, orange), for the POSEIDON retrieval (top rows) using a centered-log-ratios (CLR) agnostic prior on the background gas, and for the SCARLET retrieval (bottom rows) where H$_2$/He is the filler (background) gas. Different colors correspond to different retrievals where spots or faculae are included or excluded (see legend). Our constraints on the HMMW atmospheric volatile species are broadly unaffected by our treatment of stellar contamination, and the retrieval only provides a lower limit on the H$_2$O abundance. For the SCARLET retrieval, the distributions on $T_\mathrm{spot}$ and $T_\mathrm{fac}$ are derived from the posterior distributions on $\Delta T_\mathrm{spot}$, $\Delta T_\mathrm{fac}$ and $T_\mathrm{phot,star}$. The POSEIDON prior on the spot temperature extends all the way to 2300\,K, while SCARLET assumes that spots are at most 800\,K colder than the photosphere.}
    \label{fig:1d_distri_retrieval}
\end{figure*}

\begin{figure}
    \centering
    \includegraphics[width=0.86\textwidth]{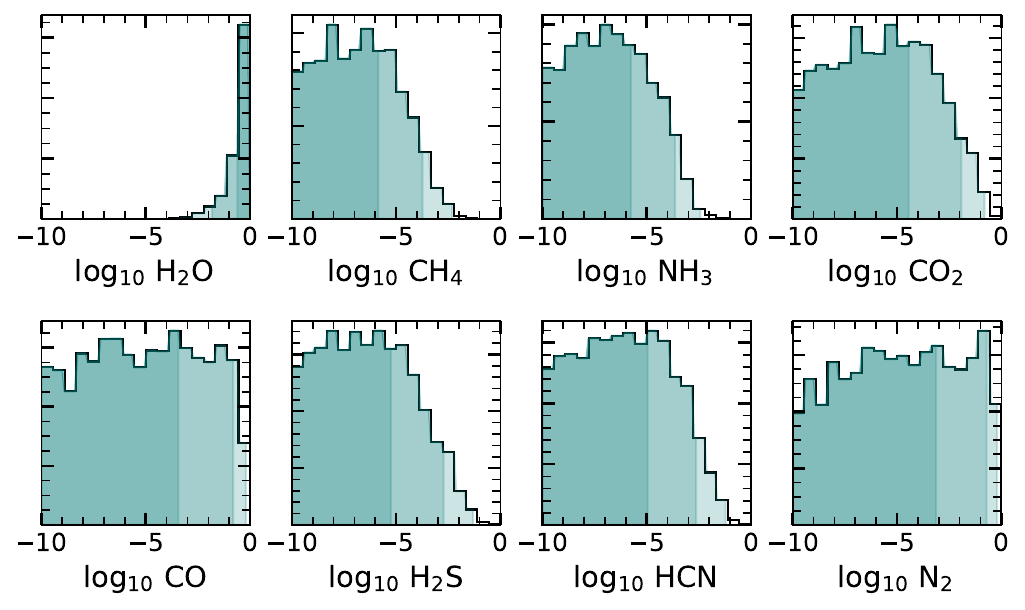}
    \vspace{5mm}
    \vspace{-5mm}\caption{Posterior distributions on the molecular abundances for all the fitted molecules in the SCARLET retrieval with H$_2$/He used as a filler gas and without stellar contamination. The 1, 2, and 3$\sigma$ lower (upper) limits for H$_2$O (other molecules) are indicated by the color shadings.}
    \label{fig:vmr_constraints_allmols}
\end{figure}

\begin{figure}
    \centering
    \includegraphics[width=0.86\textwidth]{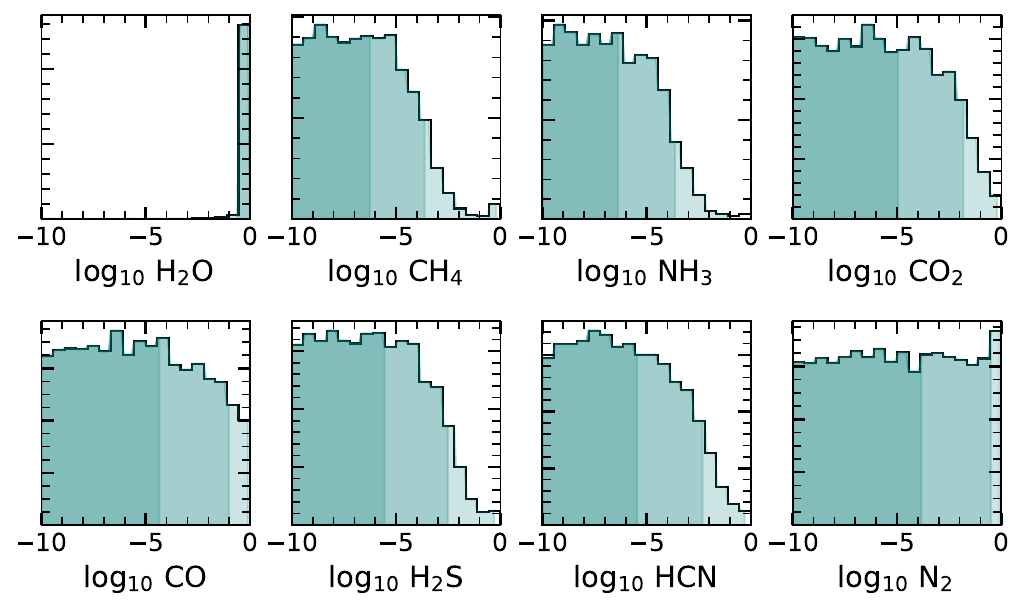}
    \vspace{5mm}
    \vspace{-5mm}\caption{Posterior distributions on the molecular abundances for all the fitted species in the POSEIDON retrieval with centered-log-ratios priors and without stellar contamination. The 1, 2, and 3$\sigma$ lower (upper) limits for H$_2$O (other molecules) are indicated by the color shadings.}
    \label{fig:vmr_constraints_allmols_clr}
\end{figure}

We use Bayesian model comparison to test the significance of the detection of stellar heterogeneities and H$_2$O in the \textit{JWST} NIRISS/SOSS + HST/WFC3 spectrum of GJ 9827 d (Table \ref{table:model comparison}). We also display the posterior distributions on H$_2$O, CH$_4$, NH$_3$ as well as stellar contamination parameters for various retrieval scenarios, with or without stellar contamination and for different assumptions about the background gas in the prior, to illustrate the robustness of our results across retrieval assumptions and frameworks (Figure \ref{fig:1d_distri_retrieval}). The posterior distributions on all the fitted molecules are also provided for the SCARLET and POSEIDON retrievals (Figures \ref{fig:vmr_constraints_allmols} and \ref{fig:vmr_constraints_allmols_clr}).

\begin{table*}
\caption{Retrieval results comparison metrics. We list model comparison statistics: the log Bayesian evidence $\ln Z$, and the $\Delta \ln Z$ for each nested model assumption compared to the baseline model without this assumption. For both the SCARLET and POSEIDON retrievals with H$_2$/He as the filler gas, the other molecules included were H$_2$O (except when explicitly removed), N$_2$, CH$_4$, NH$_3$, CO, CO$_2$, HCN, and H$_2$S. The results with the centered-log-ratios prior were obtained using POSEIDON. The `sigma' confidence levels quoted in the right-most column are calculated from the log-evidence values using the standard Bayesian comparison scale \citep{trotta_bayes_2008,benneke_how_2013}. The chemically-consistent results are quoted for SCARLET retrievals without stellar contamination. }
\centering
\begin{tabular}{lccc}
\hline
\hline
Model setup      & $\ln Z$ &  $\Delta \ln Z$  & Comment      \\

\hline
\multicolumn{3}{l}{\textbf{Impact of stellar contamination}} & \\
\hline
\textit{POSEIDON free retrievals}      &     &  &   \\

CLR, no st. contam.  & 1470.78 &  0.0& --   \\
CLR, spots  & 1471.59 &  +0.81& Weak evidence (1.90$\sigma$)  \\
CLR, spots + faculae  & 1470.55   & -1.04 &  Rejected w/ moderate confidence (2.05$\sigma$)    \\

\textit{SCARLET free retrievals}      &  &     &   \\

H$_2$/He, no st. contam.  & -942.315 &  0.0& --   \\
H$_2$/He, spots  & -942.040 &  +0.275& Weak evidence (1.45$\sigma$)  \\
H$_2$/He, spots + faculae  & -944.295   & -2.25 &  Rejected w/ moderate confidence (2.65$\sigma$)    \\

\hline

\multicolumn{3}{l}{\textbf{Impact of removing H$_2$O}} & \\

\hline
\textit{POSEIDON free retrievals}       &   &  &   \\

CLR, no st. contam., no H$_2$O  & 1467.20   &-3.58&  H$_2$O favored at 3.16$\sigma$  \\
CLR, spots, no H$_2$O  & 1469.78 &  -1.81 & H$_2$O favored at 2.45$\sigma$   \\

\textit{SCARLET free retrievals}      &   &  &   \\

H$_2$/He, no st. contam., no H$_2$O  & -949.210   &-6.895&  H$_2$O favored at 4.13$\sigma$  \\
H$_2$/He, spots, no H$_2$O  & -949.213 &  -7.170 & H$_2$O favored at 4.19$\sigma$   \\

\hline
\multicolumn{3}{l}{\textbf{Disequilibrium depletion in methane or ammonia}} & \\
\hline
\multicolumn{3}{l}{\textit{SCARLET chemically-consistent retrievals}} & \\
Chemical equilibrium  & -943.701   &0.0&  -- \\
Chemical equilibrium, CH$_4$ depl.  & -942.899   &+0.802&  Weak evidence (1.89$\sigma$)  \\
Chemical equilibrium, NH$_3$ depl.  & -943.892   &-0.191&  Rejected w/ weak confidence (1.36$\sigma$)  \\



\hline
\label{table:model comparison}
\end{tabular}
\end{table*}

\section{Internal structure modeling for GJ 9827d}

\begin{figure*}
	\centering
	{
		\includegraphics[width=0.8\textwidth]{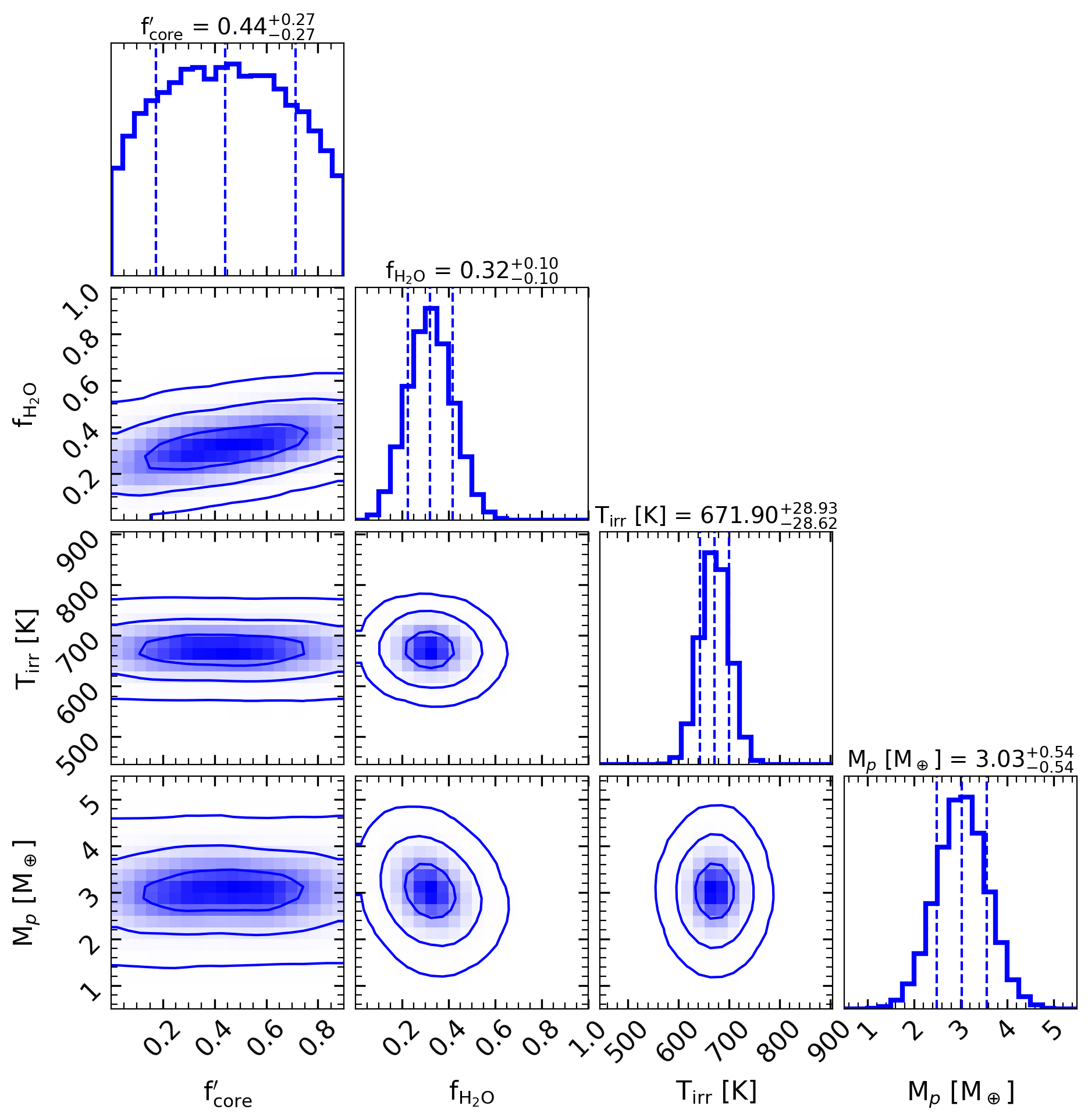}
	}
	\caption{Composition of GJ 9827 d for a 100\% water envelope. Joint and marginalized posterior distributions of the planet structure fit for GJ 9827 d assuming a vapor/supercritical hydrosphere on top of an interior composed of various mass fractions of rock and iron. The 1, 2, and 3$\sigma$ probability contours are shown.}
	\label{fig:smint_results}
\end{figure*}

We show the posterior distribution on all the parameters fitted in the Bayesian framework \texttt{smint} over the grid of water-rich models, which provide us a constraint on the water mass fraction of GJ 9827 d (Figure \ref{fig:smint_results}).

\bibliography{references, Water_worlds}

\end{document}